\documentclass{article}
\usepackage[a4paper]{geometry}
\usepackage[myheadings]{fullpage}
\usepackage{fancyhdr}
\usepackage{graphicx, wrapfig, setspace, booktabs}
\usepackage[labelfont=bf]{caption}
\usepackage{fourier}
\usepackage[english]{babel}
\usepackage{sectsty}
\usepackage{amsthm}
\usepackage{enumerate}
\usepackage{bigints}
\usepackage{bm}
\usepackage{appendix}
\usepackage{leftidx}
\usepackage[colorlinks]{hyperref}
\usepackage[numbers,sort]{natbib}
\usepackage{comment}
\usepackage{float}
\usepackage{interval}
\intervalconfig{
    soft open fences,
}

\newcommand{\normOne}[1]{\left\lVert#1\right\rVert_{1}}

\newcommand{\normF}[1]{\left\lVert#1\right\rVert_{F}}

\newcommand{\modulus}[1]{\mid{#1}\mid}

\newtheorem{theorem}{Theorem}

\newtheorem{proposition}{Proposition}
\newtheorem{definition}{Definition}
\newtheorem{assumption}{Assumption}

\newtheorem{lemma}{Lemma}
\def\XXint#1#2#3{{\setbox0=\hbox{$#1{#2#3}{\int}$}
     \vcenter{\hbox{$#2#3$}}\kern-.5\wd0}}

\usepackage[noabbrev, capitalise]{cleveref}
\crefname{axiom}{Axiom}{Axioms}
\crefname{assumption}{Assumption}{Assumptions}
\crefname{lemma}{Lemma}{Lemmas}
\crefname{proposition}{Proposition}{Propositions}
\crefname{theorem}{Theorem}{Theorems}
\crefname{table}{Table}{Tables}
\crefname{equation}{Equation}{Equations}
\crefname{hypothesis}{Hypothesis}{Hypotheses}

\numberwithin{definition}{section}
\numberwithin{lemma}{section}
\numberwithin{proposition}{section}
\numberwithin{theorem}{section}
\numberwithin{example}{section}

\newcommand{\I}{\mathbb{I}}
\newcommand{\T}{\mathbb{T}}
\newcommand{\E}{\mathbb{E}}
\newcommand{\C}{\mathbb{C}}
\newcommand{\R}{\mathbb{R}}
\newcommand{\M}{\mathbb{M}}
\newcommand{\Q}{\mathbb{Q}}
\newcommand{\X}{\mathbb{X}}

\newcommand{\Fc}{\mathcal{F}}
\newcommand{\Lc}{\mathcal{L}}

\newcommand{\Zc}{\mathcal{Z}}
\newcommand{\Rc}{\mathcal{R}}
\newcommand{\Bc}{\mathcal{B}}
\newcommand{\Xc}{\mathcal{X}}

\newcommand{\diff}{\mathrm{d}}
\newcommand{\dd}{\mathrm{d}}
\newcommand{\diag}{\mathrm{diag}}

\renewcommand{\modulus}[1]{\lvert #1 \rvert}
\newcommand{\covar}[1]{\langle #1 \rangle}

\newcommand{\matsq}[1]{\mathcal{M}_{#1}(\R)}
\newcommand{\matsqp}[1]{\mathcal{M}_{#1}(\R_+)}
\newcommand{\matsqc}[1]{\mathcal{M}_{#1}(\C)}

\newcommand{\orth}[1]{\mathcal{O}_{#1}(\R)}
\newcommand{\spd}[1]{\mathcal{S}_{#1}(\R)}
\newcommand{\pd}[1]{\mathcal{S}_{#1}^{+}(\R)}

\newcommand{\ctop}{*}

\newcommand{\Hardy}{\mathbb{H}^2}

\newcommand{\nassets}{d}

\newcommand{\avgN}{\theta}

\newcommand{\avgvolume}{v}

\newcommand{\lapl}[1]{\widehat{#1}}

\newcommand{\ciperm}{\Lambda}
\newcommand{\cikernel}{K}

\newcommand{\tcikernel}{\Gamma}

\newcommand{\mcikernel}{K^1}
\newcommand{\acikernel}{K^2}
\newcommand{\reduced}[1]{\overset{\vee}{#1}}

\newcommand{\imbkernel}{\varphi}

\title{A characterisation of cross-impact kernels \thanks{Mathieu Rosenbaum and Mehdi Tomas gratefully acknowledge financial support of the ERC 679836 Staqamof and the Chair Analytics and Models for financial regulation, Deep finance and statistics, Machine learning and systematic methods. Mehdi Tomas gratefully acknowledges the \textit{Econophysics \& Complex Systems} Research Chair under the aegis of the Fondation du Risque, a joint initiative by the \textit{Fondation de l'\'Ecole polytechnique, l'\'Ecole polytechnique} and Capital Fund Management. The authors thank Michael Benzaquen, Iacopo Mastromatteo and Michele Vodret for helpful discussions and comments.}}
\author{Mathieu Rosenbaum\footnote{CMAP, \'Ecole Polytechnique, mathieu.rosenbaum@polytechnique.edu} \and Mehdi Tomas \footnote{CMAP \& LadHyx, \'Ecole Polytechnique, mehdi.tomas@polytechnique.edu}}
\date{\today}

\begin{document}

\maketitle

\begin{abstract}
Trading a financial asset pushes its price as well as the prices of other assets, a phenomenon known as cross-impact. We consider a general class of kernel-based cross-impact models and investigate suitable parameterisations for trading purposes. We focus on kernels that guarantee that prices are martingales and anticipate future order flow (martingale-admissible kernels) and those that ensure there is no possible price manipulation (no-statistical-arbitrage-admissible kernels). We determine the overlap between these two classes and provide formulas for calibration of cross-impact kernels on data. We illustrate our results using SP500 futures data.
\end{abstract}



%%%% Introduction

\section*{Introduction}

How do trades move prices of financial securities? It is well-known among practitioners and academics that buying a financial asset tends to push its price up while selling it tends to push its price down. This observation is one aspect of \textit{market impact}, which describes how trades on one asset translate into its price. The many studies on market impact \cite{Bouchaud2018TradesPrices,Almgren2005DirectImpact,Torre1997BARRAHandbook} have deepened our understanding of how markets digest trades into prices. In turn, this has helped us understand key properties of dynamics of asset prices. For instance, market impact explains why price volatilities are well-modeled by rough fractional Brownian motions~\cite{jusselin2018no}.
\\ \\
Yet classical market impact does not tell us how our trades influence prices of other assets. Thus, it ignores a potentially important aspect of price formation. As many assets are simultaneously traded in financial markets, this element is required to generate complete market dynamics and, ultimately, answer the question of how markets digest liquidity.
\\ \\
To solve this issue, a recent strand of works \cite{Tomas2019FromModels,tomas2020build,wang2016cross,wang2017grasping,Benzaquen2017DissectingAnalysis,Alfonsi2016MultivariateFunctions,tomas2021cross,Schneider2017Cross-impactNo-dynamic-arbitrage} has studied \emph{cross-impact}, which describes how transactions on a universe of instruments drive their prices. This paper contributes to the literature by characterising cross-impact models which lead to well-behaved market dynamics. We show how these models can be calibrated from empirical data and provide an example using SP500 futures.
\\ \\
We consider a stylised market made of $\nassets$ financial securities, continuously quoted and traded by market participants. Trading activity on financial markets is highly endogenous: statistically, trades trigger other trades \cite{hardiman2013critical,Bouchaud2018TradesPrices}. To capture this effect, we model trade dynamics in our market by Hawkes processes \cite{Hawkes1971PointProcesses.} and introduce the $\nassets$-dimensional counting processes $N^a_t/N^b_t$ denoting the number of buy/sell orders of all market participants over $[0,t]$.  We classically assume that trades induce permanent impact which is linear in the traded volume, for reasons related to no-arbitrage \cite{Schneider2017Cross-impactNo-dynamic-arbitrage,Gatheral2009No-Dynamic-ArbitrageImpact}. A large class of models which satisfy this property and has been well-studied when $\nassets=1$ is the class of propagator models \cite{Bouchaud2018TradesPrices}. Thus we restrict ourselves to this class and assume that $\nassets$-dimensional price process $P$ evolves as
\begin{equation}
\label{eq:intro_propagator}
P_t = P_0 + \int_{0}^{t} \cikernel(t-s) (\dd N^{a}_s- \dd N^{b}_s) \, ,
\end{equation}
where $\cikernel \colon t \in \R_+ \mapsto \cikernel(t) \in \matsq{\nassets}$ is a \textit{cross-impact kernel}. The cross-impact kernel encodes all information about cross-impact in our market but, contrary to prices and trades, it is not directly observable. When it exists, the limit of the cross-impact kernel $ \ciperm := \underset{t \to \infty}{\lim} \cikernel(t)$ is called the permanent cross-impact matrix since $\ciperm_{ij}$ quantifies the permanent price impact of a trade on Asset $j$ on the price of Asset $i$.
\\ \\
Within \cref{eq:intro_propagator}, we examine two different classes of cross-impact kernels: those that anticipate upcoming order flow and yield martingale prices, which we dub \textit{martingale-admissible}, and those that prevent statistical arbitrage, which we dub \textit{no-statistical-arbitrage-admissible}, or \textit{nsa-admissible} for short. Statistical arbitrage is meant in the sense of~\cite{Gatheral2009No-Dynamic-ArbitrageImpact}: a statistical arbitrage is a trading strategy that starts and ends with no asset holdings and has negative expected costs.
\\ \\
Martingale-admissible and nsa-admissible kernels enforce different aspects of price efficiency. Martingale-admissible kernels ensure that prices are not predictable and that information flow is reflected in the current price, so that no trading strategy can make a profit by forecasting prices or order flows. On the other hand, nsa-admissible kernels prevent price manipulation by large agents who could push prices to make a profit. 
\\ \\
The main contribution of this paper is the characterisation of the class of martingale-admissible and nsa-admissible kernels with respect to price and order flow statistics. In particular, at most one cross-impact kernel is both martingale-admissible and nsa-admissible. This characterisation can be used for calibration on real data and we provide an application on market data to illustrate our results.
\\ \\ 
We now comment on the links between our approach and the literature.
\\ \\
This market model using Hawkes processes in a propagator framework is a generalisation of the model from~\cite{jusselin2018no,Jaisson2015MarketImbalance}, where only one asset is considered. In the single-asset case, this market dynamic is consistent with many empirical results concerning market impact. Therefore the multivariate generalisation of this framework will stay consistent with these findings while providing insights into cross-impact.
\\ \\
The papers \cite{Schneider2017Cross-impactNo-dynamic-arbitrage,Benzaquen2017DissectingAnalysis} study a class of cross-impact kernels which give rise to martingale prices. This condition is used to obtain a calibration methodology. Though the calibration methodology based on maximum likelihood is straightforward, the resulting cross-impact kernels are noisy, prone to overfitting and they have no guarantees of no-arbitrage. We show that all martingale-admissible kernels, including those presented in \cite{Schneider2017Cross-impactNo-dynamic-arbitrage,Benzaquen2017DissectingAnalysis}, have a certain form.
\\ \\
The class of nsa-admissible kernels has been described in \cite{Alfonsi2016MultivariateFunctions}. However, the provided characterisation is quite theoretical and gives no insight into which cross-impact kernels to choose in practice. This paper extends some of the results of \cite{Alfonsi2016MultivariateFunctions}. In particular, we show that nsa-admissible kernels have constrained values at zero and infinity which are related to price and order flow statistics. Unfortunately, there are many kernels which are nsa-admissible but which lead to ill-behaved market dynamics, as pointed out in \cite{Alfonsi2016MultivariateFunctions}. We provide a methodology for obtaining a nsa-admissible kernel which is close to a martingale-admissible kernel. Doing so, we obtain a kernel which is still faithful to empirical data while preventing statistical arbitrage.
\\ \\
The resulting kernels can be used on market data to estimate cross-impact. Thus, the paper adds to the literature focusing on calibrating cross-impact kernels~\cite{Hasbrouck2001CommonLiquidity,Pasquariello2015StrategicMarket,Schneider2017Cross-impactNo-dynamic-arbitrage,Wang2015PriceResults,Benzaquen2017DissectingAnalysis}. While some kernels are nsa-admissible such as the eigenliquidity cross-impact kernels \cite{Benzaquen2017DissectingAnalysis}, others are martingale-admissible \cite{Benzaquen2017DissectingAnalysis,Schneider2017Cross-impactNo-dynamic-arbitrage}. This paper provides kernels which can be easily calibrated in both classes.
\\ \\
Finally, the boundary values $\cikernel(0)$ and $\underset{t \to \infty}{\lim} \cikernel(t)$ of any cross-impact kernel $\cikernel$ which is martingale-admissible and satisfies necessary conditions for nsa-admissability have a microscopic foundation. Indeed, both can be interpreted as solutions to the multivariate version of Kyle's insider trading problem~\cite{delMolino2018TheDifferent}. Thus, although cross-impact is purely a reaction to order flow imbalance in our model, the cross-impact kernel can also be interpreted through the lens of information revelation.
\\ \\
The paper is organized as follows. In \cref{sec:setup}, we describe our financial market. In \cref{sec:cikernel}, we characterise the classes of martingale-admissible and nsa-admissible kernels. Finally, we apply our results on market data in \cref{sec:application_data} before concluding in \cref{sec:conclusion}. Some proofs and additional results are relegated to an appendix.

\section*{Notation}

The set of $\nassets \times \nassets$ real-valued square matrices is denoted by $\matsq{\nassets}$, the set of orthogonal (also called rotation) matrices by $\orth{\nassets}$, the set of real symmetric matrices by $\spd{\nassets}$ and the set of real symmetric positive matrices by $\pd{\nassets}$. Furthermore, given a matrix $A$ in $\matsq{\nassets}$, $A^\top$ denotes its transpose. Given $A$ in $\pd{\nassets}$, we write $A^{1/2}$ for a matrix such that $A^{1/2} (A^{1/2})^\top = A$ and $\sqrt{A}$ for the square root matrix, the unique positive semi-definite symmetric matrix such that $(\sqrt{A})^2 = A$. Finally, given a vector $v \in \R^{\nassets}$, we write $v = (v_1, \dots, v_{\nassets})$ and $\diag(v)$ for the diagonal matrix with entries the elements of $v$.
\\ \\
A matrix $M \in \matsq{\nassets}$ is called non-negative if for any $x \in \R^{\nassets}$, $x^\top M x \geq 0$. It is called non-negative definite if $z^* M z \geq 0$ for any $z \in \C^{\nassets}$. A matrix $M \in \matsqc{\nassets}$ is called strictly positive if $x^\top M x > 0$ for any nonzero $x \in \R^{\nassets}$ and strictly positive definite if $z^* M z > 0$ for any nonzero $z \in \C^{\nassets}$. The conjugate transpose of a matrix $M \in \matsqc{\nassets}$ is written $M^*$.
\\ \\ 
A function $f \colon \R \to \R$ is called causal if $f(t) = 0$ for all negative $t$. The same is said of a vector-valued or matrix-valued function if all its entries are causal. Given an integrable function $f$, we denote its Fourier transform $\lapl{f} \colon \C \to \C$ defined, for all $\omega \in \C$ such that the integral converges, by
$$
\lapl{f}(\omega) = \int_{-\infty}^{\infty}f(t) e^{-i\omega t} \dd t \, .
$$
Similarly, we define the Fourier transform of a vector-valued or matrix-valued function with integrable entries by the vector-valued or matrix-valued function Fourier transform of all its entries. For a given real-valued measure $\mu \colon \Bc(\R) \to \R$, we denote its Fourier transform by $\lapl{\mu} \colon \C \to \C$ defined, for all $\omega \in \C$ such that the integral converges, by
$$
\lapl{\mu}(\omega) = \int_{-\infty}^{\infty}e^{-i\omega t} \mu(\dd t) \, .
$$
Similarly, we define the Fourier transform of a vector or matrix with measure entries by the Fourier transform of all its entries. All stochastic processes in the text are defined on a probability space $(\Omega, \mathcal{F}, (\mathcal{F}_t)_{t \in \R}, \mathbb{P})$. Given two semi-martingale processes $X$ and $Y$, we denote by $\covar{X,X}$ the predictable quadratic variation of $X$ and $\covar{X,Y}$ the predictable quadratic covariation of $X$ and $Y$.

\section{Market model}
\label{sec:setup}

This section presents the stylised market model in force throughout the paper. Our setting extends~\cite{Jaisson2015MarketImbalance,jusselin2018no} to the multivariate case. For a lengthier discussion about these assumptions in the univariate case, we refer the reader to~\cite{Jaisson2015MarketImbalance,jusselin2018no}. 
\\ \\
We consider a market made of $\nassets$ different assets, quoted and traded continuously in time. We greatly simplify the market by abstracting away microstructural features and assume that agents can buy and sell Asset $i$ at time $t$ at the unique quoted price $P^i_t$. The $\nassets$-dimensional price process is denoted by $P := (P^1, \cdots, P^\nassets)$. We assume that agents trade from time $0$ onward. During this period, the cumulative traded volume by all agents at the ask (resp. bid) is denoted by $V^a$ (resp. $V^b$) and the net traded volume by $V := V^a - V^b$.
\\ \\
A key property of the order flow in financial markets is its persistence: the sign correlation of orders is slowly decaying in time \cite{Bouchaud2018TradesPrices}. Because of this effect, a particularly successful model for order flow dynamics is the Hawkes process which can capture self-excitation and cross-excitation across time and instruments \cite{Hawkes1971PointProcesses.,Bacry2015HawkesFinance}. Thus we will assume that the order flow dynamics are given by a Hawkes process.

\begin{assumption}[Hawkes order flow]
\label{ass:hawkes}
The number of buy and sell market orders follows a Hawkes process $(N^{a}, N^{b})$, of intensity $(\lambda^{a}, \lambda^{b})$ and kernel $\Phi = \begin{pmatrix} \Phi^{a/a} & \Phi^{a/b} \\
\Phi^{b/a} & \Phi^{b/b}\end{pmatrix}$ such that
\begin{align*}
    \lambda^{a}_t &= \mu + \int_{0}^{t} \Phi^{a/a}(t-s) \dd N^{a}_{s} + \int_{0}^{t} \Phi^{a/b}(t-s) \dd N^{b}_{s} \\
    \lambda^{b}_t &= \mu + \int_{0}^{t} \Phi^{b/a}(t-s) \dd N^{a}_{s} + \int_{0}^{t} \Phi^{b/b}(t-s) \dd N^{b}_{s} \, ,
\end{align*}
where in the above 
\begin{itemize}
    \item the vector $\mu \in \R_+^{\nassets}$ is the exogenous intensity of buy and sell market orders;
    \item the entry-wise integrable matrix function $\Phi^{a/a} \colon t \mapsto \Phi^{a/a}(t) \in \matsqp{\nassets}$ (resp. $\Phi^{b/b} \colon t \mapsto \Phi^{b/b}(t) \in \matsqp{\nassets}$) encodes the endogenous contribution of past buy (resp. sell) market orders on the intensity of buy (resp. sell) market orders;
    \item the entry-wise integrable matrix function $\Phi^{a/b} \colon t \mapsto \Phi^{a/b}(t) \in \matsqp{\nassets}$ (resp. $\Phi^{b/a} \colon t \mapsto \Phi^{b/a}(t) \in \matsqp{\nassets}$) encodes the endogenous contribution of past sell (resp. buy) market orders on the intensity of buy (resp. sell) market orders.
\end{itemize} 
We assume that the Hawkes parameters are such that $\E[\lambda^a_t] = \E[\lambda^b_t]$ for any $t$. Each market order on Asset $i$ is assumed to be of constant size $\avgvolume_i$ and the spectral radius of the $L_1$ norm of the Hawkes kernel $\Phi$ is assumed to be below one. The latter assumption allows us to define the stationary version of the Hawkes process (in fact stationary intensity).
\end{assumption}
This framework allows for rich multivariate dynamics since we can account for self-excitation and buy/sell interactions between different assets through $\Phi^{b/a}_{ij}$ and $\Phi^{a/b}_{ji}$. We assumed that the market there are as many buy market orders than sell market orders on the small time scales of the market model. 
\\ \\
While the order flow is persistent, the order flow at time $t$ should not give information about the order flow at time $t' \gg t$. We formalize this in the next assumption.
\begin{assumption}[Finitely predictable orderflow]
\label{ass:finitelypredictable}
For all $t \geq 0$, $\E[V^a_s - V^b_s \mid \Fc_t]$ converges in probability to some finite limit as $s$ tends to infinity.
\end{assumption}
Note that this assumption implies additional constraints on the Hawkes parameters.
\\ \\
In our market, prices are driven by the transactions of all agents. To exclude trivial arbitrages and keep the model simple, it is natural to assume that the permanent component of the cross-impact is a linear function of the order flow (see for example Corollary 3.7 of \cite{Schneider2017Cross-impactNo-dynamic-arbitrage}) and to consider a propagator framework as explained in the introduction \cite{Bouchaud2018TradesPrices}. This leads to the following assumption.
\begin{assumption}[Price dynamics]
	\label{ass:price_dynamics}
	There exists some function $\cikernel \colon t \in \R_+ \mapsto \cikernel(t) \in \matsq{\nassets}$, called a cross-impact kernel, such that the price process $P$ satisfies, for all $t \in \R_+$
	\begin{equation}
	\label{eq:propagator_assumption}
	P_t = P_0 + \int_{0}^{t} \cikernel(t-s) (\dd N^{a}_s - \dd N^{b}_s) \, ,
	\end{equation}
	and $\cikernel(t) \underset{t \to \infty}{\to} \ciperm$, where $\Lambda$ is an invertible $d \times d$ matrix called the permanent cross-impact matrix.
\end{assumption}

The diagonal functions $\cikernel_{ii}$ of the cross-impact kernel relate past order flow on a security to its price. Off-diagonal elements $\cikernel_{ij}$ relate past order flow of Asset $j$ to the price of Asset $i$. The matrix $\ciperm$ is called the permanent cross-impact matrix since $\ciperm_{ij}$ quantifies how much the price of asset $i$ is moved by the net order flow on asset $j$ after a long period.
\\ \\
Finally, we make a technical assumption about the continuity of the Hawkes kernel at the origin and its decay at infinity.

\begin{assumption}
\label{ass:diff_0}
The Hawkes order flow kernel $\Phi$ is continuously differentiable at zero and square-integrable. 
\end{assumption}

This assumption is not really constraining since Hawkes kernels for order flows are found to be square-integrable when calibrated on financial data \cite{hardiman2013critical}.
\\ \\
Previous hypotheses have not touched on the efficiency of prices in our market. Without imposing additional assumptions, prices may be highly predictable or agents could manipulate them through trading to generate profits. These two concepts of price efficiency are not always compatible, even in this stylised model. As such, we need to distinguish between cross-impact kernels which give martingale prices and those that prevent statistical arbitrage. 
\\ \\
The class of cross-impact kernels which give rise to martingale prices also includes trivial examples, such as $\cikernel = 0$. To exclude these, we introduce \textit{martingale-admissible} kernels, which anticipate the impact contribution of the order flow and lead to martingale prices. Such kernels generate non-trivial price dynamics since they incorporate the impact contribution of trades in prices. In the univariate case $d=1$, Theorem 2.1 of \cite{Jaisson2015MarketImbalance} shows that when prices are martingales and trades impact prices, we have
$$
P_t - P_0 = \kappa \underset{s \to \infty}{\lim} \E[V^a_s - V^b_s \mid \Fc_t] \, ,
$$
where $\kappa > 0$ is the permanent market impact contribution. This motivates the following definition for \textit{martingale-admissible} kernels.

\begin{definition}[Martingale-admissible kernels]
A cross-impact kernel $\cikernel$ is said to be martingale-admissible if
\begin{equation}
\label{eq:martingale_autocorr}
P_t - P_0 = \int_{0}^{t} \cikernel(t-s) (\dd N^{a}_s - \dd N^{b}_s) = \ciperm \underset{s \to \infty}{\lim} \E[V^a_s - V^b_s \mid \Fc_t] \, .
\end{equation}
\end{definition}

We later show that martingale-admissible kernels lead to martingale prices. Such kernels anticipate the market order flow to set martingale prices according to linear permanent cross-impact. This prevents agents who successfully forecast order flow to trade profitably. However, it does not forbid statistical arbitrages entirely. Before introducing relevant definitions, we define trading strategies within our market model below.

\begin{definition}[Trading strategy]
\label{def:trading_strategy}
The buy and sell trades sent under the trading strategy $f \colon \R \to \R^{\nassets}$ are $\nassets$-dimensional Poisson processes $n^a$ and $n^b$, independent of the Hawkes process $(N^a,N^b)$, with intensities given by $f^a := \max(f, 0)$ and $f^b := \max(-f, 0)$. The (average) cost of the trading strategy $f$ is
\begin{equation}
\label{eq:cost_trading_strategy}
C(f) := \int_{0}^{\infty} \int_{0}^{t} f(t)^\top K(t-s) f(s) \dd s \dd t \, .
\end{equation}
If $\int_{0}^{\infty} f(s) \dd s = 0$, the trading strategy is called a round-trip strategy and if $f$ has finite support, it is called a finite-horizon trading strategy.
\end{definition}

All trading strategies considered in this paper have deterministic intensity, ignore exchange fees, bid-ask spreads and other microstructural trading costs, so that trading costs are exclusively induced by market impact. The average cost \cref{eq:cost_trading_strategy} is derived since under the agent's trading strategy the price process becomes
$$
P_t = P_0 + \int_{0}^{t} \cikernel(t-s) (\dd N^a_s - \dd N^b_s + \dd n^a_s - \dd n^b_s) \, ,
$$
so that the average trading cost of the strategy is
\begin{align*}
\E \left[ \int_{0}^{\infty} (\dd n^a_t - \dd n^b_t)^\top (P_t - P_0) \right] &= \E \left[ \int_{0}^{\infty} (f^a(t) - f^b(t))^\top (P_t - P_0) \dd t \right] \\
&= \E \left[ \int_{0}^{\infty} \int_{0}^{t}  (f^a(t) - f^b(t))^\top \cikernel(t-s)  (\dd N^a_s - \dd N^b_s + \dd n^a_s - \dd n^b_s) \dd t \right].
\end{align*}
Therefore, since the counting processes $n^a$ and $n^b$ are independent from each other and from the Hawkes process $(N^a, N^b)$ and $\E[\lambda^a_t] = \E[\lambda^b_t]$ for all $t$, we obtain
\begin{align*}
\E \left[ \int_{0}^{\infty} (\dd n^a_t - \dd n^b_t)^\top (P_t - P_0) \right] &=  \int_{0}^{\infty} \int_{0}^{t}  (f^a(t) - f^b(t))^\top \cikernel(t-s)  (f^a(s) - f^b(s)) \dd s \dd t \\
&= \int_{0}^{\infty} \int_{0}^{t}  f(t)^\top \cikernel(t-s)  f(s) \dd s \dd t \, .
\end{align*}
This justifies \cref{eq:cost_trading_strategy}. Trading strategies which are profitable on average are called statistical arbitrages and defined below.

\begin{definition}[Statistical arbitrage]
A statistical arbitrage is a finite horizon, round-trip trading strategy such that its costs are negative:
$$
C(f) < 0 \, .
$$
\end{definition}

Cross-impact kernels which allow for statistical arbitrage induce important issues for applications. For example, they would bias trading strategies which seek to minimise trading costs towards trading-induced price manipulation. For such problems, we require trading costs models with theoretical guarantees of no statistical arbitrage. We call \textit{no-statistical-arbitrage-admissible} (or \textit{nsa-admissible} for short) cross-impact kernels that prevent statistical arbitrage.

\begin{definition}[No-statistical-arbitrage-admissible kernels]
A cross-impact kernel $\cikernel$ is said to be nsa-admissible if there are no possible statistical arbitrages, i.e. no round-trip trading strategies with average negative cost.
\end{definition}

Finally, we make the distinction between readily available information, such as prices and trades, and non-directly observable information, such as the cross-impact kernel $\cikernel$ or the permanent cross-impact matrix $\ciperm$. We refer to empirical observables for information which is easily mesaurable.

\begin{definition}[Empirical observables]
An empirical observable is a first or second-order moment measure of the price or order flows counting processes.
\end{definition}

Definition of the moment measures are given in \cref{sec:moment_measures}. Loosely speaking, they can be seen as the moments of our market variables. Empirical observables play an important role: they can be understood as key features of our stylised market, which we measure and use to derive the cross-impact kernel $\cikernel$. The next section shows that they constrain the class of relevant cross-impact kernels. 
\\ \\
Our model being set, the next section presents the main results of the paper.

\section{Characterisation of cross-impact kernels}
\label{sec:cikernel}

The previous section introduced the framework in force throughout the paper. We now characterise the cross-impact kernels $\cikernel$ which emerge from these assumptions, depending on the hypotheses on the market. We characterise martingale-admissible kernels in \cref{sec:martingale} and nsa-admissible kernels in \cref{sec:no-arbitrage}. Finally, \cref{sec:synthesis} concludes on the cross-impact kernels which are both martingale-admissible and nsa-admissible. 

\subsection{Characterisation of martingale-admissible kernels}
\label{sec:martingale}

In this section, we focus on characterising martingale-admissible kernels. We begin by characterising martingale-admissible cross-impact kernels as a function of Hawkes parameters and the permanent cross-impact matrix $\ciperm$. Then, we express these cross-impact kernels as a function of empirical observables.

\subsubsection{Cross-impact kernel as a function of Hawkes parameters}

The following proposition derives the martingale-admissible cross-impact kernels $\cikernel$ as a function of Hawkes parameters and the permanent cross-impact matrix $\ciperm$.

\begin{proposition}
\label{prop:ci_kernel_func_phi}
For any martingale-admissible kernel $\cikernel$, the price is a martingale, $\Phi^{b/b} - \Phi^{a/b} = \Phi^{a/a} - \Phi^{b/a}$ and for all $t \in \R_+$
$$
\cikernel(t) = \cikernel(0) (\I_d - \int_{0}^{t} \imbkernel(s) \dd s) \, ,
$$
where we have introduced the imbalance kernel
$$
\imbkernel := \Phi^{b/b} - \Phi^{a/b} = \Phi^{a/a} - \Phi^{b/a} \, .
$$
Furthermore, the immediate cross-impact matrix and permanent cross-impact matrix are related as follows:
$$
K(0) = \ciperm \left (\I_{\nassets} - \int_{0}^{\infty} \imbkernel(s) \dd s \right)^{-1} \diag(\avgvolume)^{-1} \, .
$$
\end{proposition}

The proof of \cref{prop:ci_kernel_func_phi} is given in \cref{app:proof_ci_kernel_func_phi}. Note that, by \cref{prop:ci_kernel_func_phi} and \cref{ass:diff_0}, any martingale-admissible cross-impact kernel $\cikernel$ is almost-everywhere differentiable and its derivative is square-integrable. \cref{prop:ci_kernel_func_phi} provides an expression for $\cikernel$ as a function of Hawkes parameters and the permanent cross-impact matrix $\ciperm$. However, the imbalance kernel $\imbkernel$ is hard to estimate and we cannot measure the permanent cross-impact matrix $\ciperm$ on real data. We would like to derive an analogous expression using solely empirical observables which can be easily measured on empirical data. This is the topic of the next section.

\subsubsection{Cross-impact kernel as a function of empirical observables}

To derive an expression for martingale-admissible cross-impact kernels, it is convenient to introduce the stationary version of the Hawkes order flow process. We write $\tilde{N}$ for the stationary version of the Hawkes process with baseline $\mu$ and kernel $\Phi$ (this process exists and is unique since the spectral radius of the $L_1$ norm of $\Phi$ is smaller than one, see e.g. \cite{bremaud1996stability}). Loosely speaking, $\tilde{N}$ describes the long-term behaviour of $N$. The stationary version allows us to define properly empirical observables but we still consider that the order flow process is given by the non-stationary process. 
\\ \\
We write $\Omega^{\tilde{N}}$ and $\Omega$  for the reduced covariance measures, defined in \cref{sec:moment_measures}, of the multivariate stationary point processes $\tilde{N}$ and $\tilde{N}^{b} - \tilde{N}^{a}$. By construction, we have $\Omega = (\I_{\nassets}, -\I_{\nassets}) \Omega^{\tilde{N}} (\I_{\nassets}, -\I_{\nassets})^\top$. 
\\ \\
Using the above notations, the following proposition relates martingale-admissible kernels to empirical observables and the boundary values of the cross-impact kernel $K(0)$ and $\underset{t \to \infty}{\lim}\cikernel(t) = \ciperm$.

\begin{proposition}
\label{prop:spectral_factor_kernel}
Any martingale-admissible cross-impact kernel $\cikernel$ satisfies, for almost all $\omega \in \R$
\begin{equation}
\label{eq:spectral_factor_kernel}
\lapl{\cikernel^{'}}(\omega) = \dfrac{1}{\sqrt{2}} \mathcal{G} \mathcal{O} \mathcal{L}(\omega)^{-1} - \cikernel(0) \, , 
\end{equation}
where $\mathcal{G}$ is any matrix such that $\mathcal{G} \mathcal{G}^\top = \ciperm  \int_{0}^{\infty} \Omega(\dd s)  \ciperm^\top$, $\mathcal{L}$ is any spectral factor of $\Omega$ (see \cref{def:spectral-factor}) and $\mathcal{O}$ is the unique rotation matrix such that
$$
\ciperm = \dfrac{1}{\sqrt{2}} \mathcal{G} \mathcal{O} \mathcal{L}(0)^{-1} \, ,
$$
where we recall that $\underset{t \to \infty}{\lim}\cikernel(t) = \ciperm$ and, by \cref{prop:ci_kernel_func_phi} and \cref{ass:diff_0}, $\cikernel$ is almost-everywhere differentiable and its derivative is square-integrable.
\end{proposition}

The proof of \cref{prop:spectral_factor_kernel} is given in \cref{app:proof_spectral_factor_kernel}. \cref{prop:spectral_factor_kernel} completely characterises the derivative of martingale-admissible cross-impact kernels as a function of quantities easily measurable on data through $\Sigma, \Omega$ and the boundary values $\cikernel(0)$ and $\ciperm$. However, these are not known a priori. Thus, for a given set of empirical observables, martingale-admissible cross-impact kernels may only differ by their boundary values.
\\ \\
The above characterisation for martingale-admissible cross-impact kernels is useful to calibrate martingale-admissible kernels and we make use of it in \cref{sec:application_data}. However, to do so, we must choose values for $\cikernel(0)$ and $\ciperm$. An outstanding question is thus that of appropriate values, which we address in the next section where we find that nsa-admissible kernels have constrained boundary values.

\subsection{Characterisation of nsa-admissible kernels}
\label{sec:no-arbitrage}

The previous section examined martingale-admissible cross-impact kernels. We found that such cross-impact kernels are completely constrained -- except at the boundaries. In this section, we focus on nsa-admissible cross-impact kernels. Contrary to the previous section, we will find that nsa-admissible cross-impact kernels are largely unconstrained, except at the boundary values. We begin by showing the latter.

\subsubsection{Constraints on the boundary values of the cross-impact kernel}
\label{sec:imm_cross_impact}

This section derives the boundary values for nsa-admissible cross-impact kernels. The following proposition characterises the immediate cross-impact matrix for any nsa-admissible kernel. We introduce  $\avgN := (\I_{\nassets} - \Phi^{a/a}) \mu + \Phi^{a/b} \mu$ which represents the stationary average of the intensity of incoming buy or sell orders.

\begin{proposition}
\label{prop:imm_impact_matrix}
For any nsa-admissible kernel $\cikernel$, we have
\begin{equation}
\label{eq:imm_impact_matrix}
\cikernel(0) = \dfrac{1}{\sqrt{2}} (\Lc_0^{-1})^\top \sqrt{\Lc_0^\top \Sigma \Lc_0} \Lc_0^{-1} \, ,
\end{equation}
where 
\begin{enumerate}[(i)]
    \itemsep0em
    \item the matrix $\Lc_0$ is any matrix such that $\Lc_0 \Lc_0^\top = \diag (\avgN_1 \avgvolume^2_1, \cdots,  \avgN_{\nassets} \avgvolume^2_{\nassets})$,
    \item the matrix $\Sigma := \underset{t \to \infty}{\lim} \E[\dd \covar{P,P}_t]$ is loosely speaking the stationary instantaneous covariance matrix of returns. The existence of this limit is shown in the proof of the proposition.
\end{enumerate}
\end{proposition}

The proof of \cref{prop:imm_impact_matrix} is given in \cref{app:proof_imm_matrix}. Note that $\cikernel(0)$ does not depend on the choice of $\Lc_0$ such that $\Lc_0 \Lc_0^\top = \diag (\avgN_1 \avgvolume^2_1, \cdots,  \avgN_{\nassets} \avgvolume^2_{\nassets})$. The matrix $\cikernel(0)$ has a microscopic interpretation. Indeed, within Kyle's insider trading model \cite{kyle1985continuous} extended to multiple assets \cite{delMolino2018TheDifferent} the market-maker adjusts his quotes according to the pricing rule $G q$, where $G$ is called the Kyle cross-impact matrix and $q = q_{IT} + q_{NT}$ is the aggregate order flow of the insider and noise traders. In this model, if the price-covariance matrix is $\Sigma$ and the noise order-flow covariance matrix $\E[(q_{NT})^\top q_{NT}]$ is $\Omega(\{0\}) = \diag (\avgN_1 \avgvolume^2_1, \cdots,  \avgN_{\nassets} \avgvolume^2_{\nassets})$, then $G = \cikernel(0)$, where $\cikernel(0)$ is given by \cref{prop:imm_impact_matrix}. In our model, $\Omega(\{0\})$ represents the instantaneous covariance matrix of order flow, which is diagonal since there are no simultaneous orders on different assets. Thus, although no agents in our model have information, $\cikernel(0)$ can be interpreted through the lens of information revelation.
\\ \\
With a quite similar proof as that of \cref{prop:imm_impact_matrix}, we can show that the permanent cross-impact matrix for any nsa-admissible kernel is symmetric non-negative.

\begin{proposition}
\label{prop:perm_impact_matrix}
For any nsa-admissible kernel $\cikernel$, the matrix $\underset{t \to \infty}{\lim}\cikernel(t) = \ciperm$ is symmetric non-negative.
\end{proposition}

The proof of \cref{prop:perm_impact_matrix} is given in \cref{app:proof_perm_matrix}. Importantly, the elements of \cref{eq:imm_impact_matrix} can be estimated quite easily on data. Thus, for any nsa-admissible kernel $\cikernel$, $\cikernel(0)$ can be expressed solely as a function of market observables. On the other hand, the permanent cross-impact matrix $\underset{t \to \infty}{\lim}\cikernel(t) = \ciperm$ is only constrained by symmetry and non-negativeness.

\subsubsection{Constraints on the Fourier transform of the cross-impact kernel}

The previous section showed that boundary values of nsa-admissible kernels are constrained and that their boundary value at zero is completely characterised. In this section, we derive necessary and sufficient conditions for nsa-admissible kernels. We begin with a result from~\cite{Alfonsi2016MultivariateFunctions}, which holds in a more general setting than the one in force in this paper, given in the lemma below.

\begin{lemma}[Theorem 2.10 of \cite{Alfonsi2016MultivariateFunctions}]
\label{lem:alfonsi}
A continuous cross-impact kernel $\cikernel$ is nsa-admissible if and only if there exists a matrix-valued non-negative definite Hermitian measure $\M$ such that for all $t \in \R$ we have
$$
\Zc(t) = \int_{\R} e^{i \gamma t} \M(\dd \gamma) \, ,
$$
where
$$
\Zc(t) := \begin{cases}
\cikernel(t) \textnormal{ if } t > 0 \\
\cikernel(0) \textnormal{ if } t = 0 \hspace{10pt} . \\
\cikernel(-t)^\top \textnormal{ if } t < 0
\end{cases}
$$
\end{lemma}

\begin{proof}
Since $\cikernel(0)$ is symmetric and $\cikernel$ is continuous this result stems from Theorem 2.10 of \cite{Alfonsi2016MultivariateFunctions}.
\end{proof}

In our framework, we can extend the previous result and prove some properties concerning the smoothness of nsa-admissible kernels, which is the topic of the next proposition.

\begin{proposition}
\label{prop:differentiability}
A continuous cross-impact kernel $\cikernel$ is nsa-admissible kernel if and only if, using the notations of \cref{lem:alfonsi}, one of the following identities holds for almost all $t \in \R$
\begin{align*}
\Zc^{'}(t) &= i \int_{\R} \gamma e^{i \gamma t} \M(\dd \gamma) \\
\Zc^{''}(t) &= - \int_{\R} \gamma^2 e^{i \gamma t} \M(\dd \gamma) \, ,
\end{align*}
where $\M$ is a matrix-valued non-negative definite Hermitian measure such that $\Zc(t) = \int_{\R} e^{i \gamma t} \M(\dd \gamma)$ and each integral converges absolutely, i.e. each integrand is absolutely integrable with respect to the measure $\M$. If any of these conditions is satisfied, the matrix function $\Zc$ has twice continously differentiable entries.
\end{proposition}

The proof of \cref{prop:differentiability} is given in \cref{app:proof_differentiability}. The regularity properties of \cref{prop:differentiability} enable us to check monotonicity and convexity of continuous nsa-admissible cross-impact kernels, a topic of interest as discussed in \cite{Alfonsi2016MultivariateFunctions}.

\subsection{Characterisation of martingale and nsa-admissible kernels}
\label{sec:synthesis}

This section summarises the results from the two previous ones to characterise kernels which are martingale-admissible and nsa-admissible. We have seen that martingale-admissible kernels are constrained everywhere except at the boundaries, while nsa-admissible kernels are constrained at the boundaries but are largely unconstrained elsewhere. It is thus natural to observe that there is at most one kernel that is both martingale-admissible and nsa-admissible. In practice, this candidate kernel is not always nsa-admissible. For applications, it may be interesting to slightly relax the martingale property in order to guarantee no statistical arbitrage. Thus, we introduce a regularisation technique to obtain a kernel close to the kernel which gives martingale prices but that prevents arbitrage. 
\\ \\
The next proposition shows that for kernels which are both nsa-admissible and martingale-admissible, the permanent cross-impact matrix is fixed. We recall that the matrix $\Sigma = \underset{t \to \infty}{\lim} \E[\dd \covar{P,P}_t]$ is loosely speaking the stationary instantaneous covariance matrix of returns.

\begin{proposition}
\label{prop:perm_impact_matrix_expression}
For any nsa-admissible, martingale-admissible kernel $\cikernel$, we have
\begin{equation}
\label{eq:perm_impact_matrix}
\underset{t \to \infty}{\lim} \cikernel(t) = \ciperm = \dfrac{1}{\sqrt{2}} (\Lc_{\infty}^{-1})^\top \sqrt{\Lc_{\infty}^\top \Sigma \Lc_{\infty}} \Lc_{\infty}^{-1} \, ,
\end{equation}
where the matrix $\Lc_{\infty}$ is any matrix such that $\Lc_{\infty} \Lc^\top_{\infty} = \int_{0}^{\infty} \Omega(\dd s)$, which is loosely speaking the stationary total autocovariance matrix of order flows.
\end{proposition}

\begin{proof}
As $\cikernel$ is both martingale-admissible and nsa-admissible, \cref{prop:perm_impact_matrix,prop:imm_impact_matrix,prop:spectral_factor_kernel} imply that $\ciperm$ is a symmetric, non-negative matrix that satisfies
$$
\ciperm \int_{0}^{\infty} \Omega(\dd s) \ciperm^\top = \dfrac{1}{2} \Sigma \, .
$$
The result follows.
\end{proof}
 
The previous proposition highlights that, as $\cikernel(0)$ in \cref{prop:imm_impact_matrix}, the permanent cross-impact matrix given in \cref{eq:perm_impact_matrix} has a microscopic interpretation. However, while $\cikernel(0)$ can be interpreted as the market-maker pricing rule in a market where the price covariance is $\Sigma$ and the noise order flow covariance matrix is $\Omega(\{0\}) = \diag (\avgN_1 \avgvolume^2_1, \cdots,  \avgN_{\nassets} \avgvolume^2_{\nassets})$, the permanent cross-impact matrix $\ciperm$ can be interpreted as the market-maker pricing rule in a market where the price covariance is $\Sigma$ and the order flow covariance is $\int_{0}^{\infty} \Omega(\dd s)$. The latter can be interpreted as the total order flow covariance which encapsulates instantaneous and non-instantaneous liquidity.
\\ \\
The next theorem summarises the results of the paper to completely characterise cross-impact kernels which are both martingale-admissible and nsa-admissible.

\begin{theorem}
There exists a unique cross-impact kernel $\cikernel$ that is martingale-admissible and which satisfies the necessary conditions for arbitrage-admissibility outlined in \cref{eq:imm_impact_matrix,eq:perm_impact_matrix}. Its expression is given by inverting \cref{eq:spectral_factor_kernel} and setting the boundary values given by \cref{eq:imm_impact_matrix,eq:perm_impact_matrix}. Furthermore, if it satisfies \cref{prop:differentiability}, then $\cikernel$ is also nsa-admissible.
\end{theorem}

\begin{proof}
Let $\cikernel$ be a cross-impact kernel which is both martingale-admissible and satisfies \cref{eq:imm_impact_matrix,eq:perm_impact_matrix}. Then by \cref{prop:ci_kernel_func_phi}, it is continuous. Furthermore, it must satisfy \cref{prop:ci_kernel_func_phi,prop:imm_impact_matrix} so that it is unique and its boundary values are given by \cref{eq:imm_impact_matrix,eq:perm_impact_matrix}. Since it is continuous, it is nsa-admissible if and only if it satisfies the necessary and sufficient conditions of \cref{prop:differentiability} (or \cref{lem:alfonsi}).
\end{proof}

The theorem shows there exists only one martingale-admissible kernel which satisfies the boundary conditions of nsa-admissible kernels. Though it may not be nsa-admissible, it is certainly closer to being nsa-admissible than other martingale-admissible kernels since it satisfies necessary conditions of nsa-admissible kernels. Given its importance, we write this cross-impact kernel $\mcikernel$ in the following. 

\begin{definition}[$\mcikernel$ kernel]
The cross-impact kernel $\mcikernel$ is the unique martingale-admissible kernel that satisfies \cref{eq:imm_impact_matrix,eq:perm_impact_matrix}.
\end{definition}

While $\mcikernel$ is a good candidate for applications, we have no guarantee that this kernel is nsa-admissible. This naturally poses issues in certain applications. For example, in portfolio optimization, a trading cost model which allows for arbitrages induces spurious round-trip strategies, as shown in \cite{Alfonsi2016MultivariateFunctions}. Thus, we introduce a regularisation method to find a kernel close to this candidate but which is nsa-admissible, which we write $\acikernel$. This motivates to the following definition.

\begin{definition}[$\acikernel$ kernel]
The cross-impact kernel $\acikernel$ is defined as
$$
\acikernel = \underset{\cikernel \mathrm{ arbitrage-admissible}}{\mathrm{argmin}} \ \normF{\lapl{\mcikernel} - \lapl{\cikernel}} \, ,
$$
where $\normF{\cdot }$ is the Frobenius norm.
\end{definition}

The kernel $\acikernel$ exists and is unique so that the previous definition is justified. Indeed, by \cref{lem:alfonsi,prop:differentiability}, any nsa-admissible kernel $\cikernel$ must be such that $\lapl{\cikernel} + \lapl{\cikernel}^\ctop$ is non-negative. Therefore
$$
\underset{\cikernel \mathrm{ arbitrage-admissible}}{\mathrm{argmin}} \ \normF{\lapl{\mcikernel} - \lapl{\cikernel}} = \underset{\lapl{\cikernel} + \lapl{\cikernel}^\ctop \geq 0}{\mathrm{argmin}} \ \normF{\lapl{\mcikernel} - \lapl{\cikernel}} \, .
$$
Thus, $\acikernel$ can be computed from $\mcikernel$ by replacing each eigenvalue $\varrho$ of $\lapl{\mcikernel} + \lapl{\mcikernel}^\ctop$ by $\max(\varrho, 0)$. Loosely speaking, $\acikernel$ trades martingale-admissibility for arbitrage-admissibility while staying close to $\mcikernel$.
\\ \\
The next section applies our results to market data to compute $\mcikernel$ and $\acikernel$.

\section{Application to financial data}
\label{sec:application_data}

\begin{figure}[H]
    \centering
    \includegraphics[width=0.6\linewidth]{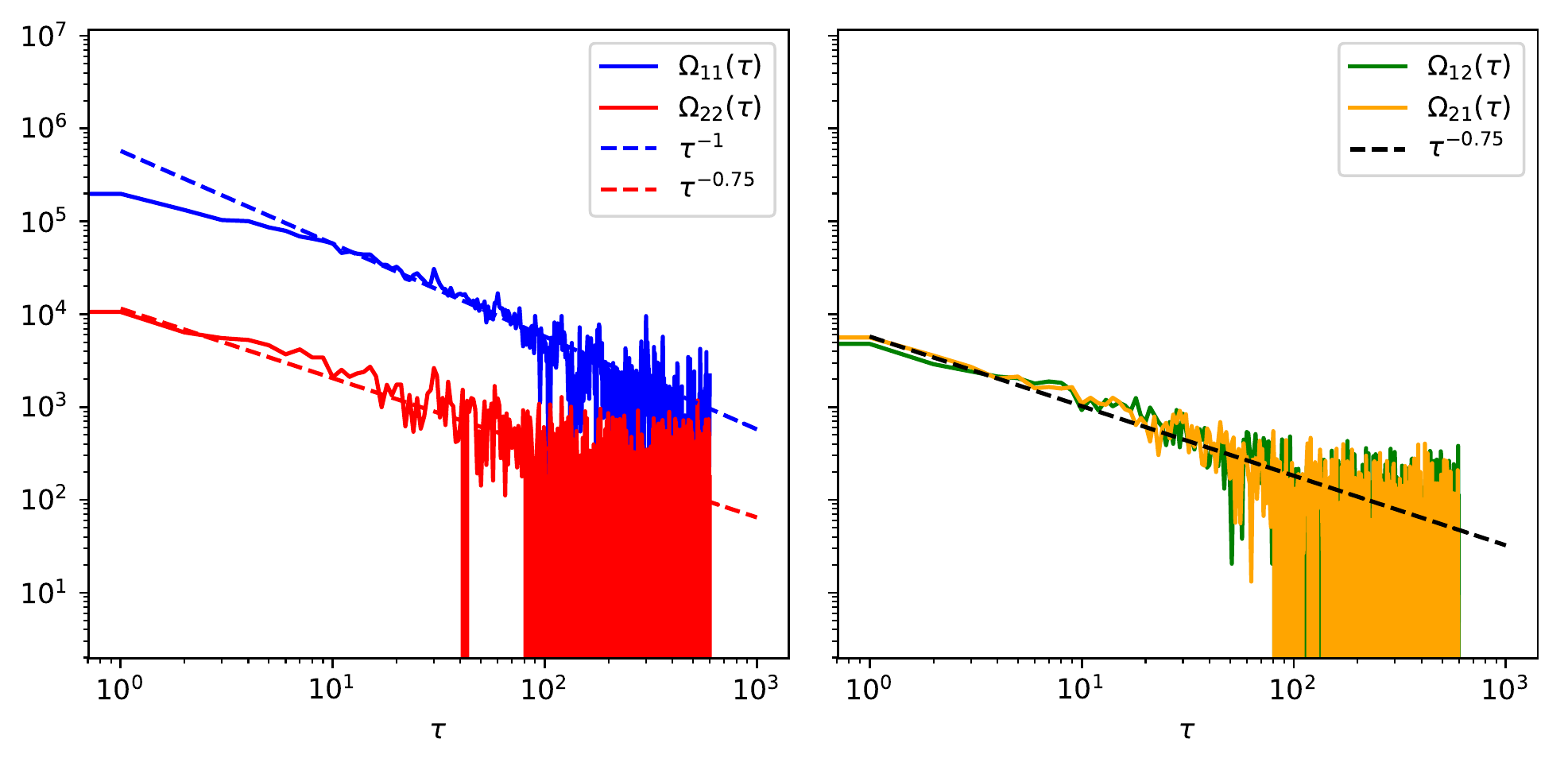}
    \caption{\textbf{Order flow auto-covariance $\Omega$.}\\
    Diagonal elements of the auto-covariance measure (left) and off-diagonal elements of the auto-covariance measure (right). Dashed lines represent power-law fits of the auto-covariances. Order flow is in product units: number of contracts traded times the contract valpoint (which is 50 for these futures). Details about the estimation procedure are given in \cref{app:estimation_empirical_observables}.}
    \label{fig:omega}
\end{figure}

This section focuses on applying the previous results to compute the kernels $\mcikernel$ and $\acikernel$ on market data. Details on the methodology and data used are given in \cref{sec:calibration_methodology}.
\\ \\
The dataset used comprises of 4 years of volumes and price data for two maturities of E-Mini SP500 futures traded on the CME, so that $\nassets = 2$. These futures are financially settled at expiry (in addition to daily settlements) according to the value of the SP500 index. The two futures selected are the lead month future, referred to as SPMINI, and the third upcoming future, referred to as SPMINI3. This data has been explored in a previous study and we refer the reader to \cite{tomas2020build} for more details into the underlying data and processing methodology.\footnote{The authors thank the \textit{Econophysics \& Complex Systems} Research Chair for providing access to this data.}
\\ \\
We begin by reporting the relevant empirical observable of our system, namely the reduced covariance measure $\Omega$ and the price-covariance matrix $\Sigma$. 
\\ \\
\cref{fig:omega} reports empirical estimates of $\Omega ( \interval[open right, scaled]{\tau \tau_s }{(\tau + 1) \tau_s} )$ for different values of $\tau$, for a time resolution of $\tau_s = 1$ second. By a slight abuse of notation, we write $\Omega(\tau)$ for $\Omega ( \interval[open right, scaled]{\tau \tau_s }{(\tau + 1) \tau_s} )$. With these conventions, $\Omega(0)$ represents the trade covariance, so that $\Omega_{11}(0)$ is the instantaneous variance of signed order flow on the front month future SPMINI and $\Omega_{22}(0)$ the instantaneous variance of signed order flow on SPMINI3. These quantities reflect the liquidity of the underlying assets since they increase with the daily traded volume \cite{tomas2020build}. The figure shows that the front-month maturity SPMINI is approximately 10 times more liquid than the SPMINI3, which highlights that most trading occurs on the leading month contract. The order flow auto-covariances $\Omega_{11}(\tau)$ and $\Omega_{22}(\tau)$ are slowly decaying in $\tau$, although $\Omega_{11}(\tau)$ exhibits faster decay than $\Omega_{22}(\tau)$. Furthermore, we observe that  $\Omega_{12}(\tau) \approx \Omega_{21}(\tau)$ so that there are no lead-lag effects in the order flows. This shows that $\Omega$ cannot be easily factorized under the form $\Omega(\tau) \approx \tau^{-\beta} C$, with some $\beta < 1$ and $C \in \matsq{\nassets}$. This hypothesis is used in certain cross-impact kernels \cite{Benzaquen2017DissectingAnalysis}.
\\ \\
The price-covariance matrix $\Sigma$, not shown here but reported in \cite{tomas2020build}, shows strong correlation ($\approx 90\%$) between the two maturities. This is natural since both futures have the same underlying.

\begin{figure}[H]
    \centering
    \includegraphics[width=0.5\linewidth]{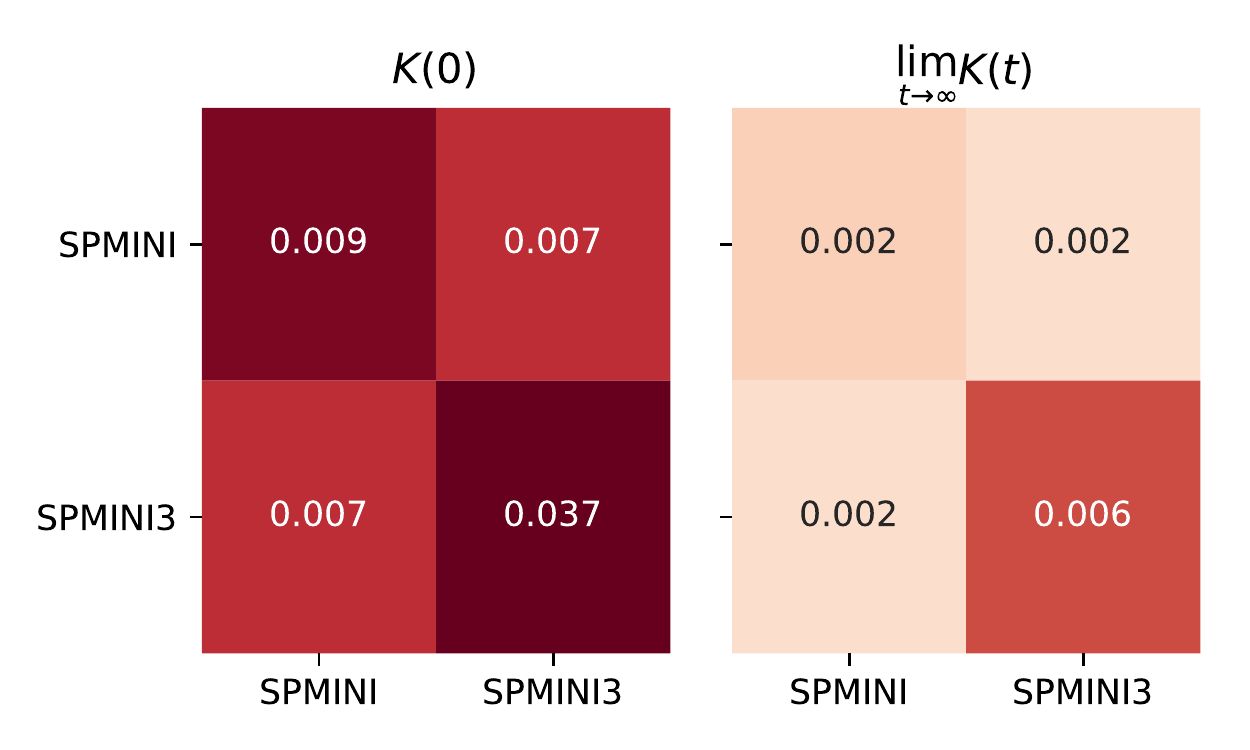}
    \caption{\textbf{Boundary values of $\mcikernel$ and $\acikernel$.}\\
    The boundary values of the martingale-admissible kernel $\mcikernel$ and the nsa-admissible kernel $\acikernel$, the immediate cross-impact matrix $\mcikernel(0) = \acikernel(0)$ and the permanent cross-impact matrix $\ciperm = \lim_{t \to \infty}\mcikernel(t) = \lim_{t \to \infty}\acikernel(t)$, estimated using \cref{prop:imm_impact_matrix} and \cref{prop:perm_impact_matrix_expression}. Values are reported in basis points (i.e. $10^{4}$ of their units). Details about the estimation procedure are given in \cref{app:computation_kernels}.}
    \label{fig:boundary}
\end{figure}

\cref{fig:boundary} shows the estimated boundary values of the $\mcikernel$ and $\acikernel$ kernels which are by definition the same for $\mcikernel$ and $\acikernel$, computed using \cref{prop:imm_impact_matrix,prop:perm_impact_matrix_expression}. The calibrated values confirm some intuitive ideas:
\begin{itemize}
	\itemsep0em
	\item price impact on the liquid future is lower than on the illiquid future: $\cikernel_{11}(0) < \cikernel_{22}(0)$ and $\ciperm_{11} < \ciperm_{22}$;
	\item buying a future immediately pushes the price of the other, as $\cikernel_{12}(0), \cikernel_{21}(0) > 0$, and the permanent impact contribution is positive since $\ciperm_{12}, \ciperm_{21} > 0$;
	\item permanent impact is lower that immediate impact, as each component of the immediate impact matrix is larger than the permanent cross-impact matrix: $\mcikernel(0) = \acikernel(0) > \ciperm$.
\end{itemize}

\begin{figure}[H]
    \centering
    \includegraphics[width=0.5\linewidth]{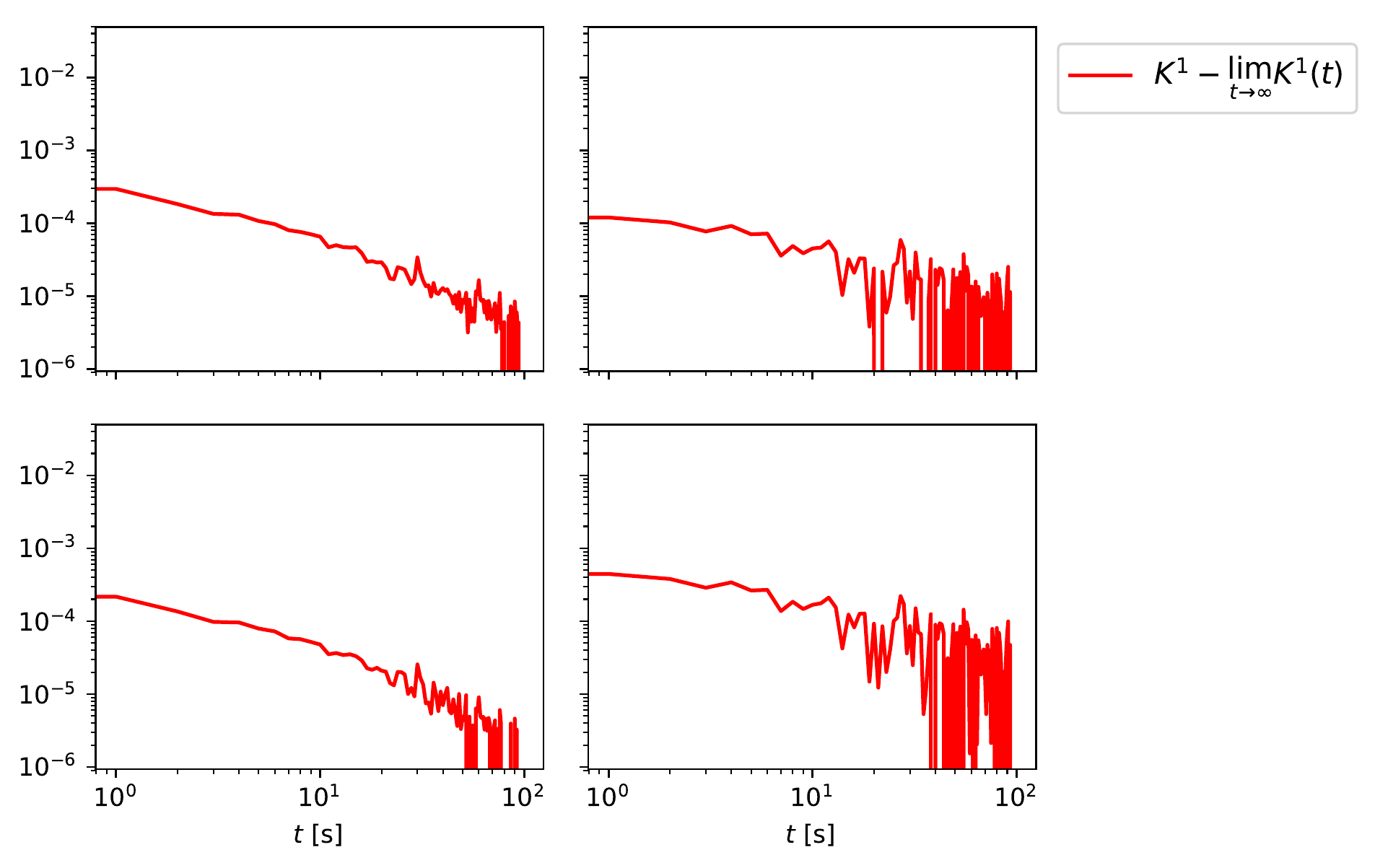}
    \caption{\textbf{Values of the $\mcikernel$ kernel.}\\
    The values of the transient part of the martingale-admissible $\mcikernel - \lim_{t \to \infty}\mcikernel(t) = \mcikernel - \ciperm$ (red) are reported. Each subplot shows the matrix elements of the kernels. For instance, the top left plot shows $\mcikernel_{11} - \ciperm_{11}$ and the top right shows $\mcikernel_{12} - \ciperm_{12}$. The permanent cross-impact matrix $\ciperm$ has been removed to highlight the power-law decay of the cross-impact kernel $\mcikernel$ toward its limit.}
    \label{fig:mci_kernel}
\end{figure}

\begin{figure}[H]
    \centering
    \includegraphics[width=0.5\linewidth]{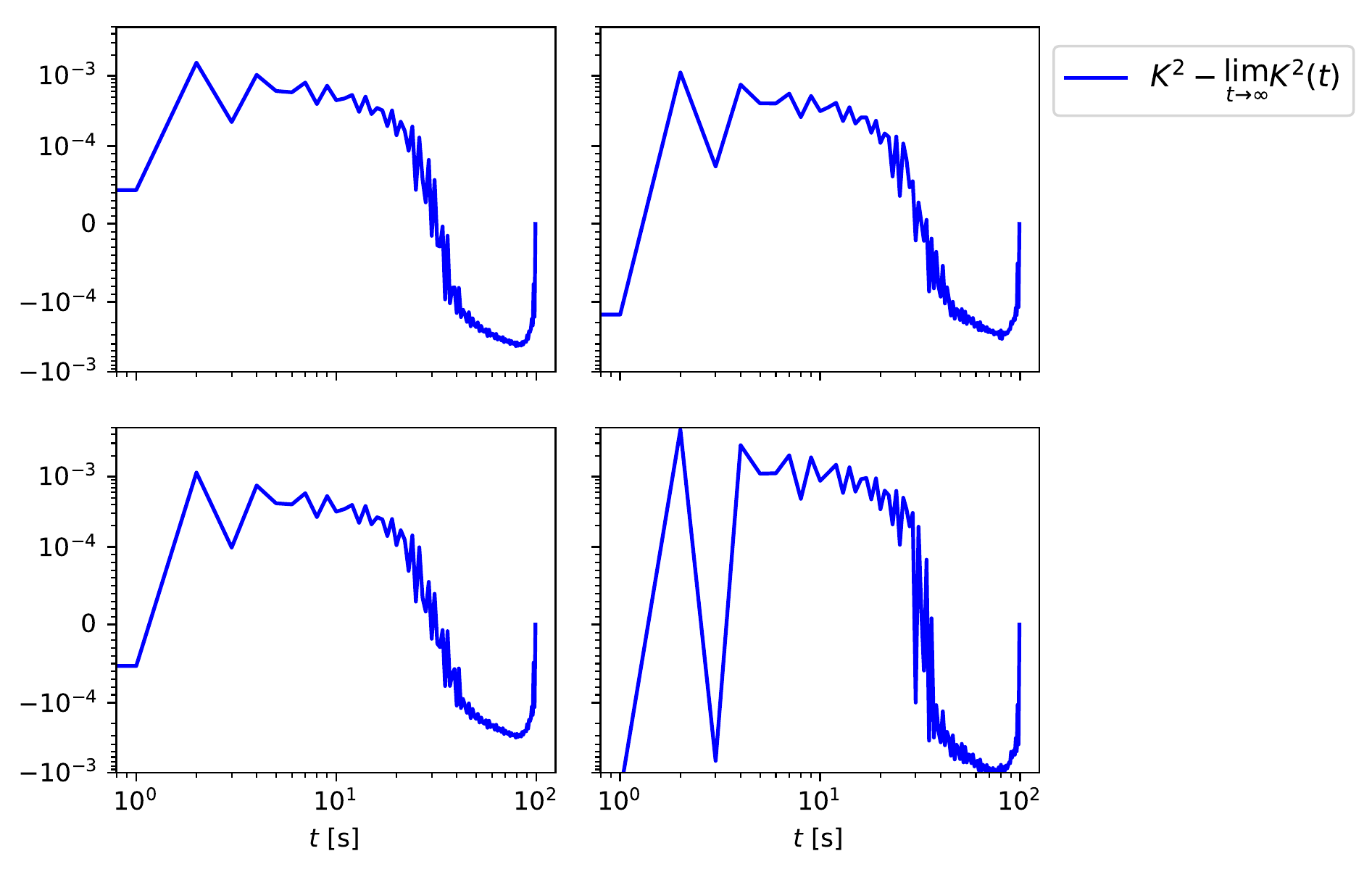}
    \caption{\textbf{Values of the $\acikernel$ kernel.}\\
    The values of the transient part of the nsa-admissible kernel $\acikernel - \lim_{t \to \infty}\acikernel(t) = \acikernel - \ciperm$ (blue) are reported. Each subplot shows the matrix elements of the kernels. For instance, the top left plot shows $\mcikernel_{11}$ and $\acikernel_{11} - \ciperm_{11}$ and the top right shows $\acikernel_{12} - \ciperm_{12}$. The permanent cross-impact matrix $\ciperm$ has been removed to highlight the behaviour of the transient part of the cross-impact kernel $\acikernel$.}
    \label{fig:aci_kernel}
\end{figure}

\begin{figure}[H]
    \centering
    \includegraphics[width=0.5\linewidth]{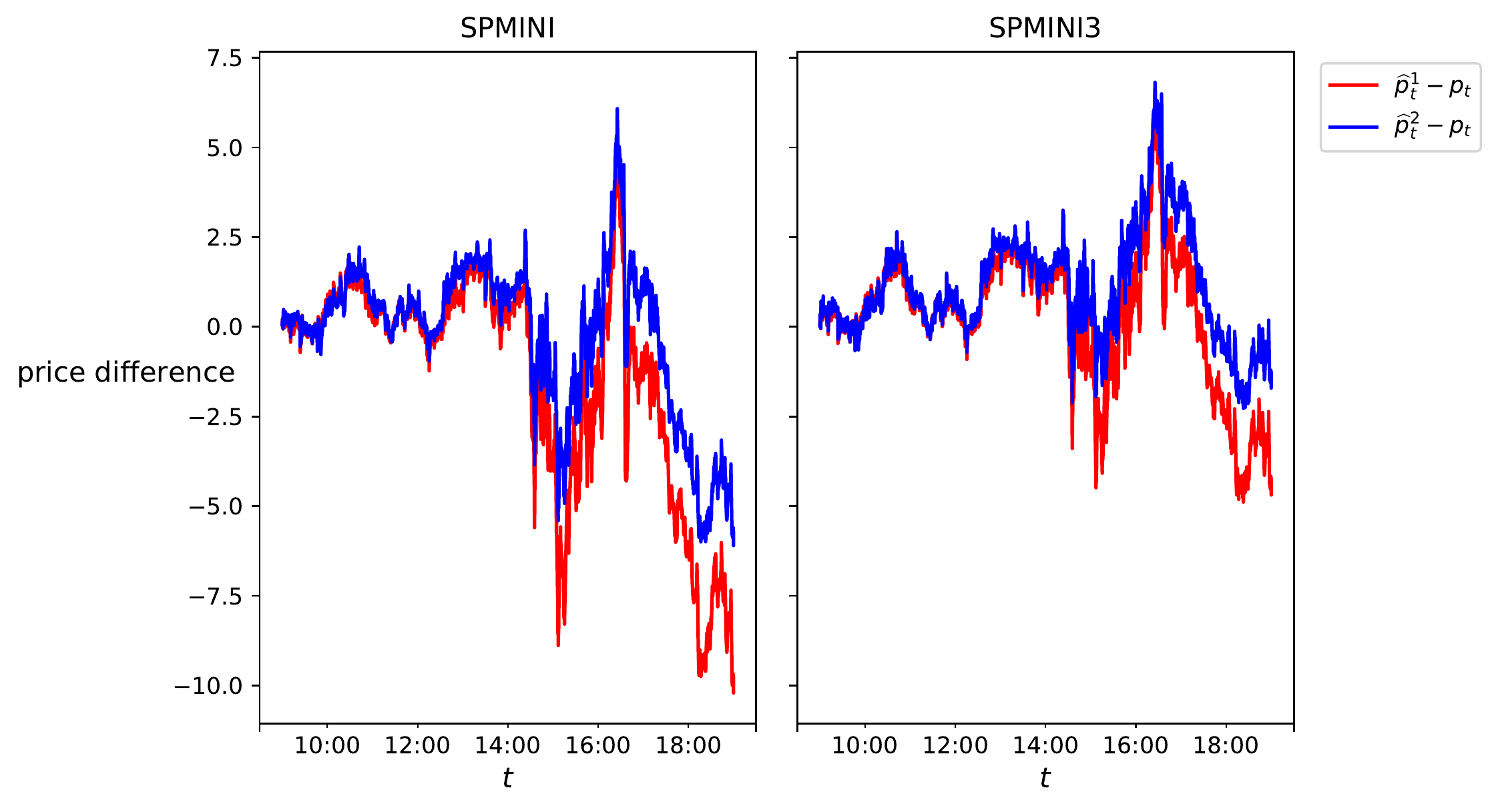}
    \caption{\textbf{Example of predicted prices from $\mcikernel$ and $\acikernel$.}\\
    The difference between predicted prices between the realized price and the impact-induced price from \cref{eq:intro_propagator} with the martingale-admissible cross-impact kernel $\mcikernel$, given by $\widehat{p}^1_t = p_0 + \sum_{s < t} \mcikernel(t-s) q_s $, (red) and with the nsa-admissible cross-impact kernel $\acikernel$, given by $\widehat{p}^2_t = p_0 + \sum_{s < t} \acikernel(t-s) q_s $, (blue), where $(q_t)$ are measured signed volumes of market orders. Price difference is reported in dollars per lot.}
    \label{fig:example_day}
\end{figure}

\cref{fig:mci_kernel,fig:aci_kernel} show both cross-impact kernels. The boundary values of $\mcikernel$ and $\acikernel$ are identical and given in \cref{fig:boundary}. As could be checked numerically, the martingale-admissible kernel $\mcikernel$ is not nsa-admissible. Thus, there is no kernel which is both martingale-admissible and nsa-admissible here. Given the very strong correlations between both assets, it is not surprising that $\acikernel_{21} \approx \acikernel_{11}$ and $\mcikernel_{21} \approx \mcikernel_{11}$: trading the first maturity pushes the price of each future by roughly the same amount. Finally, a somewhat surprising feature of $\acikernel$ is that it is non-monotonous. The cross-impact kernel $\acikernel$ is sensitive to numerical errors in the estimation methodology detailed in \cref{app:computation_kernels}, which may explain the strange value of a point of $\acikernel_{22}$. Nevertheless, it has little incidence on the kernel's fit to data (as we can see from \cref{fig:example_day}).

We illustrate the predictions of the different kernels in \cref{fig:example_day}. For a given trading day taken on the 31st of January 2017, we measure the traded order flows and build the predicted price changes according to the cross-impact rule \cref{eq:propagator_assumption}. We complete this procedure with the martingale-admissible kernel $\mcikernel$ and the nsa-admissible kernel $\acikernel$. We then compare the predicted price changes to the actual price changes. As is consistent with the literature on price impact, we see that price changes predicted from our cross-impact models are qualitatively consistent with realised price changes. However, cross-impact models are able to use trades on the liquid maturity (SPMINI) to explain price changes on the illiquid maturity (SPMINI3). This is critical since the prices of the two maturities are strongly correlated but most trades occur on the the leading month contract. Although the two kernels are quite different, their predictions are strikingly similar. This highlights that our regularisation procedure was successful in finding an nsa-admissible kernel that fits data well and prevents statistical arbitrage.

\section{Conclusion}
\label{sec:conclusion}

The goal of this paper was to characterise the class of cross-impact kernels which anticipate order flows and lead to martingale prices (\textit{martingale-admissible} kernels) and the class of cross-impact kernels which prevent statistical arbitrage (\textit{no-statistical-arbitrage-admissible} kernels). To do so, we introduce a market where trades are self-exciting and price impact is linear in the total market order flow. In this model, we derive necessary and sufficient conditions for nsa-admissible and martingale-admissible kernels. In particular, we show that only one candidate, dubbed the $\mcikernel$ kernel, could be both martingale-admissible and nsa-admissible. As there are no theoretical guarantees that the $\mcikernel$ kernel prevents arbitrage, we introduce the $\acikernel$ kernel which is close to $\mcikernel$ and prevents arbitrage. We find formulas for calibration of both kernels and apply them on SP500 futures.
\\ \\
One key result of this paper is that, given a set of market conditions, namely the asset price covariance which encodes co-movement of assets and the auto-covariance of trades which captures the way agents trade in the market, it may not be possible to conciliate the two notions of price efficiency: martingale prices and no-arbitrage. This is specific to the multi-asset case since in the single-asset case the two notions nicely co-exist \cite{jusselin2018no,Jaisson2015MarketImbalance}. Therefore, a problem we hope to address in future work is to find non-trivial market conditions where there exists a cross-impact kernel which is both nsa-admissible and martingale-admissible.
\\ \\
Finally, the results outlined in the paper have applications in market-making and trading costs estimation. For market-makers, the cross-impact kernels can be calibrated in practice to better capture adverse selection or price decay after trades. For portfolio managers, these models provide better estimate of trading costs and could be used to derive more optimal trading strategies which account for cross-impact and its decay.

\newpage

\bibliographystyle{plain}
\bibliography{bibliography.bib}

\begin{thebibliography}{10}

\bibitem{Alfonsi2016MultivariateFunctions}
Aurélien Alfonsi, Florian Kl{\"{o}}ck, and Alexander Schied.
\newblock {Multivariate transient price impact and matrix-valued positive
  definite functions}.
\newblock {\em Mathematics of operations research}, 41(3):914--934, 2016.

\bibitem{Almgren2005DirectImpact}
Robert Almgren, Chee Thum, Emmanuel Hauptmann, and Hong Li.
\newblock {Direct estimation of equity market impact}.
\newblock {\em Risk}, 18(7):58--62, 2005.

\bibitem{bacry2012non}
Emmanuel Bacry, Khalil Dayri, and Jean-Fran{\c{c}}ois Muzy.
\newblock Non-parametric kernel estimation for symmetric {H}awkes processes.
  {A}pplication to high frequency financial data.
\newblock {\em The European Physical Journal B}, 85(5):1--12, 2012.

\bibitem{Bacry2015HawkesFinance}
Emmanuel Bacry, Iacopo Mastromatteo, and Jean-François Muzy.
\newblock {Hawkes processes in finance}.
\newblock {\em Market Microstructure and Liquidity}, 1(01):1550005, 2015.

\bibitem{Benzaquen2017DissectingAnalysis}
Michael Benzaquen, Iacopo Mastromatteo, Zoltan Eisler, and Jean-Philippe
  Bouchaud.
\newblock {Dissecting cross-impact on stock markets: An empirical analysis}.
\newblock {\em Journal of Statistical Mechanics: Theory and Experiment},
  2017(2):23406, 2017.

\bibitem{Bouchaud2018TradesPrices}
Jean-Philippe Bouchaud, Julius Bonart, Jonathan Donier, and Martin Gould.
\newblock {\em {Trades, Quotes and Prices}}.
\newblock Cambridge University Press, 2018.

\bibitem{bremaud1996stability}
Pierre Br{\'e}maud and Laurent Massouli{\'e}.
\newblock Stability of nonlinear {H}awkes processes.
\newblock {\em The Annals of Probability}, pages 1563--1588, 1996.

\bibitem{daley2008introduction}
Daryl~J Daley and David Vere-Jones.
\newblock {\em An introduction to the theory of point processes: volume II:
  general theory and structure}.
\newblock Springer New York, 2008.

\bibitem{delMolino2018TheDifferent}
Luis~Carlos Garcia~del Molino, Iacopo Mastromatteo, Michael Benzaquen, and
  Jean-Philippe Bouchaud.
\newblock The multivariate {K}yle model: More is different.
\newblock {\em SIAM Journal on Financial Mathematics}, 11(2):327--357, 2020.

\bibitem{Gatheral2009No-Dynamic-ArbitrageImpact}
Jim Gatheral.
\newblock No-dynamic-arbitrage and market impact.
\newblock {\em Quantitative Finance}, 10(7):749--759, 2010.

\bibitem{hardiman2013critical}
Stephen~J Hardiman, Nicolas Bercot, and Jean-Philippe Bouchaud.
\newblock Critical reflexivity in financial markets: a hawkes process analysis.
\newblock {\em The European Physical Journal B}, 86(10):1--9, 2013.

\bibitem{Hasbrouck2001CommonLiquidity}
Joel Hasbrouck and Duane~J Seppi.
\newblock {Common factors in prices, order flows, and liquidity}.
\newblock {\em Journal of financial Economics}, 59(3):383--411, 2001.

\bibitem{Hawkes1971PointProcesses.}
Alan.~G Hawkes.
\newblock {Point spectra of some mutually exciting point processes.}
\newblock {\em Journal of the Royal Statistical Society: Series B
  (Methodological)}, 33(3):438--443, 1971.

\bibitem{Jaisson2015MarketImbalance}
Thibault Jaisson.
\newblock {Market impact as anticipation of the order flow imbalance}.
\newblock {\em Quantitative Finance}, 15(7):1123--1135, 2015.

\bibitem{jusselin2018no}
Paul Jusselin and Mathieu Rosenbaum.
\newblock No-arbitrage implies power-law market impact and rough volatility.
\newblock {\em Mathematical Finance}, 2018.

\bibitem{kyle1985continuous}
Albert~S Kyle.
\newblock Continuous auctions and insider trading.
\newblock {\em Econometrica: Journal of the Econometric Society}, pages
  1315--1335, 1985.

\bibitem{lukacs1970characteristic}
Eugene Lukacs.
\newblock {\em Characteristic functions}.
\newblock Griffin, 1970.

\bibitem{mcwhirter2007evd}
John~G McWhirter, Paul~D Baxter, Tom Cooper, Soydan Redif, and Joanne Foster.
\newblock An {EVD} algorithm for para-hermitian polynomial matrices.
\newblock {\em IEEE Transactions on Signal Processing}, 55(5):2158--2169, 2007.

\bibitem{Pasquariello2015StrategicMarket}
Paolo Pasquariello and Clara Vega.
\newblock {Strategic cross-trading in the US stock market}.
\newblock {\em Review of Finance}, 19(1):229--282, 2015.

\bibitem{Schneider2017Cross-impactNo-dynamic-arbitrage}
M.~Schneider and F.~Lillo.
\newblock Cross-impact and no-dynamic-arbitrage.
\newblock {\em Quantitative Finance}, 19(1):137--154, 2019.

\bibitem{titchmarsh1948introduction}
Edward~C Titchmarsh.
\newblock {\em Introduction to the theory of {F}ourier integrals}.
\newblock Clarendon Press, 1948.

\bibitem{tomas2020build}
Mehdi Tomas, Iacopo Mastromatteo, and Michael Benzaquen.
\newblock How to build a cross-impact model from first principles: Theoretical
  requirements and empirical results.
\newblock {\em arXiv preprint arXiv:2004.01624}, 2020.

\bibitem{tomas2021cross}
Mehdi Tomas, Iacopo Mastromatteo, and Michael Benzaquen.
\newblock Cross impact in derivative markets.
\newblock {\em arXiv preprint arXiv:2102.02834}, 2021.

\bibitem{Tomas2019FromModels}
Mehdi Tomas and Mathieu Rosenbaum.
\newblock {From microscopic price dynamics to multidimensional rough volatility
  models}.
\newblock {\em Advances in Applied Probability, to appear}, 2019.

\bibitem{Torre1997BARRAHandbook}
Nicolo Torre.
\newblock {BARRA market Impact model handbook}.
\newblock {\em BARRA Inc., Berkeley}, 1997.

\bibitem{python:ref}
Guido Van~Rossum and Fred~L. Drake.
\newblock {\em Python 3 Reference Manual}.
\newblock CreateSpace, Scotts Valley, CA, 2009.

\bibitem{2020SciPy-NMeth}
Pauli Virtanen, Ralf Gommers, Travis~E. Oliphant, Matt Haberland, Tyler Reddy,
  David Cournapeau, Evgeni Burovski, Pearu Peterson, Warren Weckesser, Jonathan
  Bright, St{\'e}fan~J. {van der Walt}, Matthew Brett, Joshua Wilson, K.~Jarrod
  Millman, Nikolay Mayorov, Andrew R.~J. Nelson, Eric Jones, Robert Kern, Eric
  Larson, C~J Carey, {\.I}lhan Polat, Yu~Feng, Eric~W. Moore, Jake
  {VanderPlas}, Denis Laxalde, Josef Perktold, Robert Cimrman, Ian Henriksen,
  E.~A. Quintero, Charles~R. Harris, Anne~M. Archibald, Ant{\^o}nio~H. Ribeiro,
  Fabian Pedregosa, Paul {van Mulbregt}, and {SciPy 1.0 Contributors}.
\newblock {{SciPy} 1.0: Fundamental Algorithms for Scientific Computing in
  Python}.
\newblock {\em Nature Methods}, 17:261--272, 2020.

\bibitem{wang2017grasping}
Shanshan Wang, Sebastian Neus{\"{u}}{\ss}, and Thomas Guhr.
\newblock {Grasping asymmetric information in market impacts}.
\newblock {\em arXiv preprint arXiv:1710.07959}, 2017.

\bibitem{Wang2015PriceResults}
Shanshan Wang, Rudi Sch{\"{a}}fer, and Thomas Guhr.
\newblock {Price response in correlated financial markets: empirical results}.
\newblock {\em arXiv preprint arXiv:1510.03205}, 2015.

\bibitem{wang2016cross}
Shanshan Wang, Rudi Sch{\"a}fer, and Thomas Guhr.
\newblock Cross-response in correlated financial markets: individual stocks.
\newblock {\em The European Physical Journal B}, 89(4):105, 2016.

\bibitem{wang2015multichannel}
Zeliang Wang, John~G McWhirter, and Stephan Weiss.
\newblock Multichannel spectral factorization algorithm using polynomial matrix
  eigenvalue decomposition.
\newblock In {\em 2015 49th Asilomar conference on signals, systems and
  computers}, pages 1714--1718. IEEE, 2015.

\bibitem{wiener1959factorization}
Norbert Wiener and E.J. Akutowicz.
\newblock A factorization of positive hermitian matrices.
\newblock {\em Journal of Mathematics and Mechanics}, pages 111--120, 1959.

\bibitem{wiener1957prediction}
Norbert Wiener and Pesi Masani.
\newblock The prediction theory of multivariate stochastic processes.
\newblock {\em Acta Mathematica}, 98(1-4):111--150, 1957.

\end{thebibliography}

\newpage

\appendix

\section{Moment measures for point processes}
\label{sec:moment_measures}

This section presents definitions of moment measures for point processes. Throughout this section, $\X$ denotes some Borelian subset of $\R^{\nassets}$ and $\Xc = \Bc(\X)$.

\begin{definition}[$n$-th moment measure]
	Given a univariate stationary point process $N$ on $\X$, the first and second order moment measure of the point process $N$ are measures on $\X$ and $\X^2$ defined for all $A, B \in \Xc$ as
	\begin{align*}
	M_{1}(A) &= \E[N(A)] \\
	M_{2}(A \times B) &= \E[N(A) N(B)]
	\end{align*}
	whenever these expectations exist.
\end{definition}

A key quantity is the reduced measure second order measure which we introduce below.

\begin{definition}[Reduced measure (Proposition 12.6.III of \cite{daley2008introduction})]
	There exist a reduced measure, noted $\reduced{M_{2}}$, such that for any bounded measurable function $f$ on $\X^2$ we have
	\begin{align*}
	\int_{\X} f(x_1, x_2) M_{2}(\dd x_1, \dd x_2) &=  \int_\X \int_{\X} f(x, x+y) \, , \reduced{M}_{2}(\dd y) \dd x
	\end{align*}
\end{definition}

The existence result of the reduced measure satisfying the above is shown in Proposition 12.6.III of \cite{daley2008introduction}. Of particular interest to us are the second-order reduced covariance measure introduced below.

\begin{definition}[Reduced covariance measure]
	The reduced covariance measure of a stationary point process is defined as 
	$$
	\reduced{C}_2(\dd u) = \reduced{M}_2(\dd u) - m^2 l(\dd u) \, ,
	$$
	where $l$ is the Lebesgue measure, $m$ is the mean intensity i.e the non-negative constant such that $M_1(\dd u) = m l(\dd u)$.
\end{definition}

Extensions of the above concepts to multivariate point processes are straightforward. In particular we have the following definition.

\begin{definition}[second order auto-moment measure]
	Given a $k$-dimensional, stationary point process $N$ on $\X$, the second order auto-moment measure of the point process $N$ is a measure on $\X^{2}$ defined for all $A,B \in \Xc$ and $1 \leq i,j \leq k$ as
	\begin{equation*}
	M_{ij}(A \times B) = \E[N_{i}(A) N_{j}(B)] \, ,
	\end{equation*}
	whenever this expectation exists.
\end{definition}

\begin{definition}[reduced covariance measure]
	The reduced covariance measure of a $k$-dimensional, stationary point process is defined, for all $1 \leq i,j \leq k$ as 
	$$
	\reduced{C}_{ij}(\dd u) = \reduced{M}_{ij}(\dd u) - m_i m_j l(\dd u) \, ,
	$$
	where $l$ is the Lebesgue measure, $m_i$ is the mean intensity i.e the non-negative constant such that $M_i(\dd u) = m_i l(\dd u)$.
\end{definition}

\section{Technical results}
\label{sec:spectral_factorization}

This section presents some key technical results necessary for the proofs. We begin by introducing an important functional space for our results, the Hardy space $\Hardy$.

\begin{definition}[Hardy space $\Hardy$]
The Hardy space $\Hardy$ is the space of functions $F \colon \C \to \C$ such that the following conditions are satisfied:
\begin{enumerate}[(i)]
    \item the function $F$ is holomorphic in the upper half of the complex plane,
    \item there exists a constant $C > 0$ such that, for all $\xi > 0$
    $$
    \int_{-\infty}^{\infty} \lvert F(\omega + i \xi) \rvert ^2 \dd \omega < C \, ,
    $$
    \item for almost all $\omega \in \R$, we have
    $$
    \lim_{\xi \to 0} F(\omega + i \xi) = F(\omega) \, .
    $$
\end{enumerate} 
\end{definition}

Functions in the Hardy space $\Hardy$ are closely related to the Fourier transform of causal functions as shown by the so-called Titchmarsh theorem below.

\begin{theorem}[Titchmarsh theorem~\cite{titchmarsh1948introduction}]
\label{thm:titchmarsh}
Given a real, complex-valued square-integrable function $F$, the following conditions are equivalent:
\begin{enumerate}[(i)]
    \itemsep0em
    \item the inverse Fourier transform of $F$ is a causal function,
    \item the function $F$ belongs to the Hardy space $\Hardy$.
\end{enumerate}
\end{theorem}

A key theorem used in the paper to derive the form of martingale-admissible cross-impact kernels is the matrix spectral factorization theorem.

\begin{theorem}[Matrix spectral factorization theorem \cite{wiener1957prediction,wiener1959factorization}]
\label{thm:mat-spectral-factor}
Let $F \colon \C \to \matsq{\nassets}$ be a matrix function such that $F$ is positive definite almost everywhere on the unit circle $\T := \{ z \in \C \colon \lvert z \rvert = 1 \} $ with integrable entries on the unit circle such that the Paley-Wiener condition 
\begin{equation}
    \label{eq:paley-wiener}
    \log \det F \in L_1(\T)
\end{equation}
is satisfied. Then $F$ admits a factorization on $\T$
\begin{equation}
    \label{eq:spectral-factorization}
    F(z) = F_+(z) F_+(z)^\ctop \, ,
\end{equation}
where $F_+$ is an analytic function with entries in the Hardy space $\Hardy$. Furthermore, the spectral factor $F_+$ is unique up to right multiplication by a unitary matrix (i.e. a matrix $U$ such that $U U^\ctop = \I_{\nassets}$).
\end{theorem}

As spectral factors are referenced throughout the paper, we introduce for convenience a definition below.

\begin{definition}[Spectral factor]
\label{def:spectral-factor}
Let $F \colon \C \to \matsq{\nassets}$ be a matrix function which satisfies the hypotheses of \cref{thm:mat-spectral-factor}. Then, a spectral factor $F_+$ of $F$ is an analytic function $F_+$ with entries in the Hardy space $\Hardy$ which satisfies \cref{eq:spectral-factorization}.
\end{definition}

\section{Proofs}

\subsection{Proof of \cref{prop:ci_kernel_func_phi}}
\label{app:proof_ci_kernel_func_phi}

The proof is adapted from \cite{Jaisson2015MarketImbalance}. Throughout this section we assume that there exists a martingale-admissible kernel $\cikernel$ such that the price satisfies
$$
P_t = P_0 + \int_{0}^{t} \cikernel(t-s) \dd (N^a_s-N^b_s) = P_0 + \ciperm \lim_{s \to \infty} \E[V^a_s - V^b_s \mid \Fc_t] \, .
$$
To compute the right hand side term, we use a classical result on Hawkes processes \cite{Bacry2015HawkesFinance,Jaisson2015MarketImbalance,jusselin2018no}: for all $t \geq 0$, we have
\begin{equation}
\label{eq:classical_result_intensity}
\begin{pmatrix} \lambda^a_t \\ \lambda^b_t \end{pmatrix} = \begin{pmatrix} \mu \\ \mu \end{pmatrix} + \int_{0}^{t} \Psi(t-s) \begin{pmatrix} \mu \\ \mu \end{pmatrix} \dd s + \int_{0}^{t} \Psi(t-s) \dd M_s \, ,
\end{equation}
where $\Psi := \sum_{n \geq 1} \Phi^{*n}$, $\Phi^{*n}$ being the $n$-th convolution product of the matrix function $\Phi$ and $M$ is a martingale. The next lemma shows that the compuation of the conditional expectation of the Hawkes process reduces to the compuation of the conditional expectation of the intensity.

\begin{lemma}
\label{lem:conditional_N_to_intensity}
For all $t,s \geq 0$ such that $s \geq t$, we have
$$
\E[N^a_s - N^b_s \mid \Fc_t] = N^a_t - N^b_t + \int_{t}^{s} \E[\lambda^a_u - \lambda^b_u \mid \Fc_t]  \dd u \, .
$$
\end{lemma}

\begin{proof}
Using the martingale decomposition of the Hawkes process, we have $N_t = M_t + \int_{0}^{t} \lambda_s \dd s$ where $M$ is a martingale. Therefore, 
$$
N^a_s - N^b_s = M^a_s - M^b_s + \int_{0}^{t} (\lambda^a_u - \lambda^b_u) \dd u + \int_{t}^{s} (\lambda^a_u - \lambda^b_u) \dd u \, .
$$
Using the martingale property, we obtain
$$
\E[N^a_s - N^b_s \mid \Fc_t] = N^a_t - N^b_t + \int_{t}^{s} \E[\lambda^a_u - \lambda^b_u \mid \Fc_t]  \dd u \, .
$$
\end{proof}

The next lemma computes the conditional expectation of the intensity for our Hawkes processes. It generalises Proposition 3.2 of \cite{Jaisson2015MarketImbalance}.

\begin{lemma}
\label{lem:intensity_conditional}
For all $t, s \geq 0$ such that $s \geq t$, we have
\begin{align*}
    \int_{t}^{s} \E[ \lambda^a_u - \lambda^b_u \mid \Fc_t] \dd u =& \int_{0}^{t} \int_{t-r}^{s-r} \Xi(u) \dd u (\dd N^a_r - \dd N^b_r) \\
& - \int_{0}^{t} \int_{t-r}^{s-r} \int_{0}^{t-r} \Xi(u-x) \Phi^{a/a}(x) \dd x \dd u \dd N^a_r -\int_{0}^{t} \int_{t-r}^{s-r} \int_{0}^{t-r}  \Xi(u-x) \Phi^{b/a}(x)  \dd x \dd u \dd N^a_r \\
& + \int_{0}^{t} \int_{t-r}^{s-r} \int_{0}^{t-r} \Xi(u-x) \Phi^{a/b}(x)  \dd x \dd u \dd N^b_r + \int_{0}^{t} \int_{t-r}^{s-r} \int_{0}^{t-r} \Xi(u-x) \Phi^{b/b}(x)  \dd x \dd u \dd N^b_r \, ,
\end{align*}
where $\Xi := \Psi^{a/a} - \Psi^{b/a} = \Psi^{b/b} - \Psi^{a/b}$.
\end{lemma}

\begin{proof}
We use \cref{eq:classical_result_intensity} to derive
\begin{align*}
\int_{t}^{s} \E[ \lambda^a_u \mid \Fc_t] \dd u =& \E \left[ \int_{t}^{s} \int_{0}^{u} \Psi^{a/a}(u-x) \dd M^a_x \dd u \mid \Fc_t \right] + \E \left[\int_{t}^{s} \int_{0}^{u} \Psi^{a/b}(u-x) \dd M^b_x \dd u \mid \Fc_t \right]  \, .
\end{align*}
Using the fact that $M$ is a martingale, we have
\begin{align*}
\E \left[\int_{t}^{s} \int_{0}^{u} \Psi^{a/a}(u-x) \dd M^a_x \dd u \mid \Fc_t \right] =& \int_{t}^{s} \int_{0}^{t} \Psi^{a/a}(u-x) \dd M^a_x \dd u \\
=& \int_{t}^{s} \int_{0}^{t}\Psi^{a/a}(u-x) \left( \dd N^a_x - (\mu + \int_{0}^{t} \Phi^{a/a}(x-r) \dd N^a_r + \int_{0}^{t} \Phi^{a/b}(x-r) \dd N^b_r) \dd x \right) \dd u \\
=& \int_{t}^{s} \int_{0}^{t} \Psi^{a/a}(u-x) \dd N^a_x \dd u  - \int_{t}^{s} \int_{0}^{t} \Psi^{a/a}(u-x) \mu \dd x \dd u \\
& - \int_{t}^{s} \int_{0}^{t} \Psi^{a/a}(u-x) \int_{0}^{t} \Phi^{a/a}(x-r) \dd N^a_r \dd x \dd u \\
& - \int_{t}^{s} \int_{0}^{t} \Psi^{a/a}(u-x) \int_{0}^{t} \Phi^{a/b}(x-r) \dd N^b_r \dd x \dd u \, .
\end{align*}
Consequently, 
\begin{align*}
\int_{t}^{s} \E[ \lambda^a_u \mid \Fc_t] \dd u  =& \int_{t}^{s} \int_{0}^{t} \Psi^{a/a}(u-x) \dd N^a_x \dd u + \int_{t}^{s} \int_{0}^{t} \Psi^{a/b}(u-x) \dd N^b_x \dd u \\
& - \int_{t}^{s} \int_{0}^{t} (\Psi^{a/a}+\Psi^{a/b})(u-x) \mu \dd x \dd u \\
& - \int_{t}^{s} \int_{0}^{t} \Psi^{a/a}(u-x) \int_{0}^{t} \Phi^{a/a}(x-r) \dd N^a_r \dd x \dd u - \int_{t}^{s} \int_{0}^{t} \Psi^{a/b}(u-x) \int_{0}^{t} \Phi^{b/a}(x-r) \dd N^a_r \dd x \dd u \\
& - \int_{t}^{s} \int_{0}^{t} \Psi^{a/a}(u-x) \int_{0}^{t} \Phi^{a/b}(x-r) \dd N^b_r \dd x \dd u - \int_{t}^{s} \int_{0}^{t} \Psi^{a/b}(u-x) \int_{0}^{t} \Phi^{b/b}(x-r) \dd N^b_r \dd x \dd u \, .
\end{align*}
Regrouping terms from $\lambda^a$ and $\lambda^b$ we obtain
\begin{align*}
    \int_{t}^{s} \E[ \lambda^a_u - \lambda^b_u \mid \Fc_t] \dd u =& \int_{t}^{s} \int_{0}^{t} (\Psi^{a/a}-\Psi^{b/a})(u-x) \dd N^a_x \dd u + \int_{t}^{s} \int_{0}^{t} (\Psi^{a/b}-\Psi^{b/b})(u-x) \dd N^b_x \dd u \\
& - \int_{t}^{s} \int_{0}^{t} (\Psi^{a/a} + \Psi^{a/b} - \Psi^{b/a} - \Psi^{b/b})(u-x) \mu \dd x \dd u \\
& - \int_{t}^{s} \int_{0}^{t} (\Psi^{a/a}-\Psi^{b/a})(u-x) \int_{0}^{t} \Phi^{a/a}(x-r) \dd N^a_r \dd x \dd u \\
& - \int_{t}^{s} \int_{0}^{t} (\Psi^{a/b}-\Psi^{b/b})(u-x) \int_{0}^{t} \Phi^{b/a}(x-r) \dd N^a_r \dd x \dd u \\
& - \int_{t}^{s} \int_{0}^{t} (\Psi^{a/a}-\Psi^{b/a})(u-x) \int_{0}^{t} \Phi^{a/b}(x-r) \dd N^b_r \dd x \dd u \\
& - \int_{t}^{s} \int_{0}^{t} (\Psi^{a/b}-\Psi^{b/b})(u-x) \int_{0}^{t} \Phi^{b/b}(x-r) \dd N^b_r \dd x \dd u \, .
\end{align*}
By \cref{ass:finitelypredictable}, the above must converge to a finite limit as $s$ tends to infinity. Therefore, as $\mu \neq 0$ and all elements of the matrix function $\Phi$ are non-negative we must have for all $t \geq 0$ that $\Psi^{a/a}(t) + \Psi^{a/b}(t) = \Psi^{b/a}(t) + \Psi^{b/b}(t)$. Therefore, writing $\Xi := \Psi^{a/a} - \Psi^{b/a} = \Psi^{b/b} - \Psi^{a/b}$, we have
\begin{align*}
    \int_{t}^{s} \E[ \lambda^a_u - \lambda^b_u \mid \Fc_t] \dd u =& \int_{t}^{s} \int_{0}^{t} \Xi(u-x) (\dd N^a_x - \dd N^b_x) \dd u \\
& - \int_{t}^{s} \int_{0}^{t} \Xi(u-x) \int_{0}^{t} \Phi^{a/a}(x-r) \dd N^a_r \dd x \dd u - \int_{t}^{s} \int_{0}^{t} \Xi(u-x) \int_{0}^{t} \Phi^{b/a}(x-r) \dd N^a_r \dd x \dd u \\
& + \int_{t}^{s} \int_{0}^{t} \Xi(u-x) \int_{0}^{t} \Phi^{a/b}(x-r) \dd N^b_r \dd x \dd u + \int_{t}^{s} \int_{0}^{t} \Xi(u-x) \int_{0}^{t} \Phi^{b/b}(x-r) \dd N^b_r \dd x \dd u \, .
\end{align*}
Using the change of variables $u' = u - r$ and $x' = x - r$ we obtain
\begin{align*}
    \int_{t}^{s} \E[ \lambda^a_u - \lambda^b_u \mid \Fc_t] \dd u =& \int_{0}^{t} \int_{t-x}^{s-x} \Xi(u') \dd u' (\dd N^a_x - \dd N^b_x) \\
& - \int_{0}^{t} \int_{t-r}^{s-r} \int_{0}^{t-r} \Xi(u'-x') \Phi^{a/a}(x') \dd x' \dd u' \dd N^a_r -\int_{0}^{t} \int_{t-r}^{s-r} \int_{0}^{t-r}  \Xi(u'-x') \Phi^{b/a}(x')  \dd x' \dd u' \dd N^a_r \\
& + \int_{0}^{t} \int_{t-r}^{s-r} \int_{0}^{t-r} \Xi(u'-x') \Phi^{a/b}(x')  \dd x' \dd u' \dd N^b_r + \int_{0}^{t} \int_{t-r}^{s-r} \int_{0}^{t-r} \Xi(u'-x') \Phi^{b/b}(x')  \dd x' \dd u' \dd N^b_r \, .
\end{align*}
\end{proof}

The next lemma uses the previous results to compute the conditional expectation of the counting process.

\begin{lemma}
\label{lem:conditional_counting_not_limit}
We have for all $t, s \geq 0$ such that $s \geq t$
\begin{align*}
\E[N^a_s - N^b_s \mid \Fc_t] =& N^a_t - N^b_t + \int_{0}^{t} \int_{t-r}^{s-r} \Xi(u) \dd u (\dd N^a_r - \dd N^b_r) \\
& - \int_{0}^{t} \int_{t-r}^{s-r} \int_{0}^{t-r} \Xi(u-x) \Phi^{a/a}(x) \dd x \dd u \dd N^a_r -\int_{0}^{t} \int_{t-r}^{s-r} \int_{0}^{t-r}  \Xi(u-x) \Phi^{b/a}(x)  \dd x \dd u \dd N^a_r \\
& + \int_{0}^{t} \int_{t-r}^{s-r} \int_{0}^{t-r} \Xi(u-x) \Phi^{a/b}(x)  \dd x \dd u \dd N^b_r + \int_{0}^{t} \int_{t-r}^{s-r} \int_{0}^{t-r} \Xi(u-x) \Phi^{b/b}(x)  \dd x \dd u \dd N^b_r \, ,
\end{align*}
\end{lemma}

\begin{proof}
Applying \cref{lem:conditional_N_to_intensity} along with \cref{lem:intensity_conditional}, the result is straightforward.
\end{proof}

The next lemma derives the limit of the conditional expectation of the process $N^a_s-N^b_s$.

\begin{lemma}
\label{lem:intensity_limit}
For all $t \geq 0$, we have
\begin{align*}
\lim_{s \to \infty} \E[N^a_s - N^b_s \mid \Fc_t] &= N^a_t - N^b_t + \int_{0}^{t} \left( \int_{t-x}^{\infty} \Xi(u) \dd u \right) (\dd N^a_x - \dd N^b_x)  \\
& - \int_{0}^{t} \left( \int_{t-x}^{\infty} \int_{0}^{t-x} \Xi(u-v) (\Phi^{a/a}+\Phi^{b/a})(v) \dd v \dd u \right) \dd N^a_x  \\
& + \int_{0}^{t} \left( \int_{t-x}^{\infty} \int_{0}^{t-x} \Xi(u-v) (\Phi^{a/b}+\Phi^{b/b})(v) \dd v \dd u \right) \dd N^b_x \, ,
\end{align*}
where $\Xi := \Psi^{a/a} - \Psi^{b/a} = \Psi^{b/b} - \Psi^{a/b}$.
\end{lemma}

\begin{proof}
Taking the limit as $s$ tends to infinity of \cref{lem:intensity_conditional} we obtain the almost sure convergence of $\int_{t}^{s} \E[ \lambda^a_u - \lambda^b_u \mid \Fc_t] \dd u$ towards
\begin{align*}
    \int_{t}^{\infty} \E[ \lambda^a_u - \lambda^b_u \mid \Fc_t] \dd u =&  \int_{0}^{t} \left( \int_{t-x}^{\infty} \Xi(u) \dd u \right) (\dd N^a_x - \dd N^b_x)  \\
& - \int_{0}^{t} \left( \int_{t-x}^{\infty} \int_{0}^{t-x} \Xi(u-v) (\Phi^{a/a}+\Phi^{b/a})(v) \dd v \dd u \right) \dd N^a_x  \\
& + \int_{0}^{t} \left( \int_{t-x}^{\infty} \int_{0}^{t-x} \Xi(u-v) (\Phi^{a/b}+\Phi^{b/b})(v) \dd v \dd u \right) \dd N^b_x \, ,
\end{align*}
where we have used that $\Xi$ is integrable. Finally, using \cref{lem:conditional_N_to_intensity} with the above yields the result.
\end{proof}

In order to show that the price is a martingale, we establish some elementary results about the convergence of the conditional expectations. 

\begin{lemma}
\label{lem:dominated_convergence_conditions}
For all $t, s \geq 0$ such that $s \geq t$, we write $Z^t_s :=  \E[N^a_s - N^b_s \mid \Fc_t]$ and $Z^t := \lim_{s \to \infty} \E[N^a_s - N^b_s \mid \Fc_t]$. Then, we have that
\begin{enumerate}[(i)]
    \item the random variable $Z^t$ is integrable,
    \item the random variables $(Z^t_s)$ converge almost surely to $Z^t$,
    \item there exists an integrable random variable $Y^t$ such that for all $s \geq t$, $\lvert Z^t_s \rvert \leq Y^t$.
\end{enumerate}
\end{lemma}

\begin{proof}
By \cref{lem:conditional_counting_not_limit}, we have for all $t, s \geq 0$ such that $s \geq t$
\begin{align*}
Z_s^t = \E[N^a_s - N^b_s \mid \Fc_t] =& N^a_t - N^b_t + \int_{0}^{t} \int_{t-r}^{s-r} \Xi(u) \dd u (\dd N^a_r - \dd N^b_r) \\
& - \int_{0}^{t} \int_{t-r}^{s-r} \int_{0}^{t-r} \Xi(u-x) \Phi^{a/a}(x) \dd x \dd u \dd N^a_r -\int_{0}^{t} \int_{t-r}^{s-r} \int_{0}^{t-r}  \Xi(u-x) \Phi^{b/a}(x)  \dd x \dd u \dd N^a_r \\
& + \int_{0}^{t} \int_{t-r}^{s-r} \int_{0}^{t-r} \Xi(u-x) \Phi^{a/b}(x)  \dd x \dd u \dd N^b_r + \int_{0}^{t} \int_{t-r}^{s-r} \int_{0}^{t-r} \Xi(u-x) \Phi^{b/b}(x)  \dd x \dd u \dd N^b_r \, .
\end{align*}
Therefore, writing $\lvert A \rvert$ for the matrix of absolute values of all entries of $A$, we have
\begin{align*}
    \lvert Z_s^t \rvert  \leq & N^a_t + N^b_t + \int_{0}^{t} \int_{t-r}^{s-r} \lvert \Xi(u) \rvert \dd u (\dd N^a_r + \dd N^b_r) \\
& + \int_{0}^{t} \int_{t-r}^{s-r} \int_{0}^{t-r} \lvert \Xi(u-x) \Phi^{a/a}(x) \rvert \dd x \dd u \dd N^a_r + \int_{0}^{t} \int_{t-r}^{s-r} \int_{0}^{t-r}  \lvert \Xi(u-x) \Phi^{b/a}(x) \rvert  \dd x \dd u \dd N^a_r \\
& + \int_{0}^{t} \int_{t-r}^{s-r} \int_{0}^{t-r} \lvert \Xi(u-x) \Phi^{a/b}(x) \rvert  \dd x \dd u \dd N^b_r + \int_{0}^{t} \int_{t-r}^{s-r} \int_{0}^{t-r} \lvert \Xi(u-x) \Phi^{b/b}(x) \rvert \dd x \dd u \dd N^b_r \\
 \leq & N^a_t + N^b_t + \int_{0}^{t} \int_{t-r}^{\infty} \lvert \Xi(u) \rvert \dd u (\dd N^a_r + \dd N^b_r) \\
& + \int_{0}^{t} \int_{t-r}^{\infty} \int_{0}^{t-r} \lvert \Xi(u-x) \Phi^{a/a}(x) \rvert \dd x \dd u \dd N^a_r + \int_{0}^{t} \int_{t-r}^{\infty} \int_{0}^{t-r}  \lvert \Xi(u-x) \Phi^{b/a}(x) \rvert  \dd x \dd u \dd N^a_r \\
& + \int_{0}^{t} \int_{t-r}^{\infty} \int_{0}^{t-r} \lvert \Xi(u-x) \Phi^{a/b}(x) \rvert  \dd x \dd u \dd N^b_r + \int_{0}^{t} \int_{t-r}^{\infty} \int_{0}^{t-r} \lvert \Xi(u-x) \Phi^{b/b}(x) \rvert \dd x \dd u \dd N^b_r \, .
\end{align*}
Writing $Y^t$ for the last random variable, $Y^t$ is independent from $s$, positive and we can easily check that it is of finite expectation since $\Xi$ is integrable. This proves (iii). Furthermore, we also have $\lvert Z^t \rvert \leq Y^t$ so that $Z^t$ is integrable and (i) is verified. Finally, \cref{lem:intensity_limit} proved point (ii). 
\end{proof}

\begin{lemma}
\label{lem:martingale_price}
The price is a martingale.
\end{lemma}

\begin{proof}
Since, for all $t, \tau \geq 0$ such that $\tau \leq t$ we have  
\begin{align*}
\E[P_t \mid \Fc_\tau] =& P_0 + \ciperm \diag{(\avgvolume)} \E[\lim_{s \to \infty} \E[N^a_s - N^b_s \mid \Fc_t] \mid \Fc_\tau] \\
=& P_0 +  \ciperm \diag{(\avgvolume)} \lim_{s \to \infty} \E[\E[N^a_s - N^b_s \mid \Fc_t] \mid \Fc_\tau] \\
=& P_0 +  \ciperm \diag{(\avgvolume)} \lim_{s \to \infty} \E[N^a_s - N^b_s \mid \Fc_\tau] \\
=& P_\tau \, ,
\end{align*}
where the second equality is an application of the dominated convergence theorem with the random variables $Z^t_s :=  \E[N^a_s - N^b_s \mid \Fc_t]$ and $Z^t := \lim_{s \to \infty} \E[N^a_s - N^b_s \mid \Fc_t]$, the conditions being met by \cref{lem:dominated_convergence_conditions}.
\end{proof}

We can simplify the expression of the price process using martingale-admissibility. This is the topic of the next lemma.

\begin{lemma}
The price process is of the form
$$
P_t = P_0 + \ciperm \diag(\avgvolume) \int_{0}^{t} \varsigma(t-s) (\dd N^a_s - \dd N^b_s) \, ,
$$
where $\varsigma(t) = \I_{\nassets} + \int_{t-x}^{\infty} \Xi(u) \dd u + \int_{t-v}^{\infty} \int_{0}^{t-v} \Xi(u-v) (\Phi^{a/a}+\Phi^{b/a})(v) \dd v \dd u$. Furthermore, $\cikernel$ is almost everywhere differentiable.
\end{lemma}

\begin{proof}
By assumption, we know there exists some cross-impact kernel $\cikernel$ such that for all $t \geq 0$
$$
P_t = P_0 + \int_{0}^{t} \cikernel(t-s) (\dd N^a_s - \dd N^b_s) \, .
$$
Therefore \cref{lem:intensity_limit} implies that for all $t \geq 0$
$$
\int_{t-v}^{\infty} \int_{0}^{t-v} \Xi(u-v) (\Phi^{a/a}+\Phi^{b/a})(v) \dd v \dd u = \int_{t-v}^{\infty} \int_{0}^{t-v} \Xi(u-v) (\Phi^{a/b}+\Phi^{b/b})(v) \dd v \dd u \, .
$$
And \cref{lem:intensity_limit} yields
$$
P_t = P_0 + \ciperm \diag(\avgvolume) \int_{0}^{t} \varsigma(t-s) (\dd N^a_s - \dd N^b_s) \, ,
$$
with $\varsigma(t) = \I_{\nassets} + \int_{t-x}^{\infty} \Xi(u) \dd u + \int_{t-v}^{\infty} \int_{0}^{t-v} \Xi(u-v) (\Phi^{a/a}+\Phi^{b/a})(v) \dd v \dd u$. In particular, we must have $\ciperm \diag(\avgvolume) \varsigma = \cikernel$ almost everywhere and $\cikernel$ is almost everywhere differentiable.
\end{proof}

To obtain a simpler expression for $\cikernel$ and the other results of the proposition, we adapt the proof of Proposition 3.3 of \cite{Jaisson2015MarketImbalance} to our structure of Hawkes processes. At time $t \geq 0$, either:
\begin{itemize}
    \item there are no market orders and the price is differentiable with $P^{'}_t = \int_{0}^{t} \cikernel^{'}(t-s) \dd (N^{a} - N^{b})_s $;
    \item there is a buy market order on some asset, say Asset $i$, which happens with intensity $\lambda^{b,i}_t$ and yields a price jump of $\cikernel(0) \avgvolume_i e_i$;
    \item there is a sell market order on some asset, say Asset $i$, which happens with intensity $\lambda^{b,i}_t$ and yields a price jump of $-\cikernel(0) \avgvolume_i e_i$.
\end{itemize}
Therefore, we have for all $t \geq 0$
\begin{align*}
\lim_{h \to 0} \dfrac{\E[P_{t+h} \mid \Fc_t] - P_t}{h} =&  \int_{0}^{t} \cikernel^{'}(t-s) (\dd N^{a}_s - \dd N^{b}_s) + \sum_{i=1}^{\nassets} \cikernel(0) \avgvolume_i e_i (\lambda^{a,i}_t - \lambda^{b,i}_t) \\
=&  \int_{0}^{t} \cikernel^{'}(t-s) (\dd N^{a}_s - \dd N^{b}_s) + \cikernel(0) \diag(\avgvolume) (\lambda^{a}_t - \lambda^{b}_t) \\
=& \int_{0}^{t} \cikernel^{'}(t-s) (\dd N^{a}_s - \dd N^{b}_s) - \cikernel(0) \diag(\avgvolume) \int_{0}^{t} (\Phi^{b/b} - \Phi^{a/b})(t-s) \dd N^{b}_{s} \\
& + \cikernel(0) \diag(\avgvolume) \int_{0}^{t} (\Phi^{a/a}-\Phi^{b/a})(t-s) \dd N^{a}_{s} \\
=& 0,
\end{align*}
where the last equality holds since the price is a martingale by \cref{lem:martingale_price}. Thus, we must have for all $t \geq 0$
\begin{align*}
    \cikernel^{'}(t) &= - \cikernel(0) \diag(\avgvolume) (\Phi^{b/b}(t) - \Phi^{a/b}(t)) \\
    \cikernel^{'}(t) &= - \cikernel(0) \diag(\avgvolume) (\Phi^{a/a}(t) - \Phi^{b/a}(t)) \, .
\end{align*}
To conclude, we need to derive the expression of $\cikernel(0)$. Since $\Phi^{b/b}-\Phi^{a/a}$ is integrable, $\cikernel$ converges to a finite limit and by definition of the permanent cross-impact matrix: $\underset{t \to \infty}{\lim} \cikernel(t) = \ciperm$. Thus
$$
\lim_{t \to \infty} \cikernel(t) = \cikernel(0) \diag(\avgvolume) \left( \I_{\nassets} - \int_{0}^{\infty} (\Phi^{b/b}(s) - \Phi^{a/b}(s)) \dd s \right) = \ciperm \, .
$$
Since by \cref{ass:hawkes} we have $\rho(\int_{0}^{\infty} \Phi(s) \dd s) < 1$, we get
$$
\cikernel(0) = \ciperm \left (\I_{\nassets} - \int_{0}^{\infty} (\Phi^{b/b}(s) - \Phi^{a/b}(s)) \dd s \right)^{-1} \diag(\avgvolume)^{-1} \, .
$$
Since the matrix $\ciperm$ is non-singular by \cref{ass:price_dynamics} so is the matrix $\cikernel(0)$ and we must also have
$$
\Phi^{b/b} - \Phi^{a/b} = \Phi^{a/a} - \Phi^{b/a} \, .
$$
This completes the proof of \cref{prop:ci_kernel_func_phi}.

\subsection{Proof of \cref{prop:spectral_factor_kernel}}
\label{app:proof_spectral_factor_kernel}

Because of the autocovariance structure of Hawkes processes, this section makes extensive use of Fourier transforms of measures. We recall from the notation section that the Fourier transform of a measure $\kappa$ at $\omega \in \C$ is defined, when the integral converges, as
$$
\lapl{\kappa}(\omega) = \int_{-\infty}^{\infty} e^{-i\omega t} \kappa(\dd t)\, .
$$
Writing $\avgN := (\I_{\nassets} - \Phi^{a/a}) \mu + \Phi^{a/b} \mu$ for the average of the intensity of the stationary version of the Hawkes process we know from \cite{bacry2012non} that
$$
\lapl{\Omega}^{\tilde{N}}(\omega) = (\I_{\nassets} - \lapl{\Phi}(\omega))^{-1} \diag(\avgN) ((\I_{\nassets} - \lapl{\Phi}(-\omega))^{-1})^{\top} \, .
$$
In particular, the Fourier transform of the reduced covariance measure $\Omega$ has integrable entries.
\\ \\
We begin the proof of \cref{prop:spectral_factor_kernel} with a useful lemma.

\begin{lemma}
\label{lem:consistency_K}
For all $\omega \in \R$, we have
$$
(\lapl{\cikernel^{'}}(\omega) + \cikernel(0)) \lapl{\Omega}(\omega) (\lapl{\cikernel^{'}}(\omega) + \cikernel(0))^{\ctop} = \ciperm \Omega_{\infty} \ciperm^\top \, .
$$
\end{lemma}

\begin{proof}
Writing $\avgN := (\I_{\nassets} - \Phi^{a/a}) \mu + \Phi^{a/b} \mu$ for the average of the intensity of the stationary version of the Hawkes process we know from \cite{bacry2012non} that
$$
\lapl{\Omega}^{\tilde{N}}(\omega) = (\I - \lapl{\Phi}(\omega))^{-1} \diag(\avgN) ((\I - \lapl{\Phi}(-\omega))^{-1})^{\top} \, .
$$
Therefore we have
\begin{align*}
\lapl{\Omega}^{\tilde{N}}(\omega) &= (\I_d - \lapl{\Phi}(\omega))^{-1} \diag(\avgN) ((\I_d - \lapl{\Phi}(\omega))^{-1})^{\ctop} \\
\lapl{\Omega}(\omega) &= (\I_{\nassets}, -\I_{\nassets}) \lapl{\Omega}^{\tilde{N}}(\omega) (\I_{\nassets}, -\I_{\nassets})^\top \, .
\end{align*}
Using pseudo-inverses we obtain
\begin{align*}
\lapl{\Omega}(\omega)^{-1} &= \dfrac{1}{4} (\I_{\nassets}, -\I_{\nassets}) (\lapl{\Omega}^{\tilde{N}}(\omega))^{-1} (\I_{\nassets}, -\I_{\nassets})^\top \\
 &= (\I_{\nassets} - \lapl{\imbkernel}(\omega))^\ctop \diag(\avgN)^{-1} (\I_{\nassets} - \lapl{\imbkernel}(\omega)) \\
&= (\I_d - \lapl{\imbkernel}(\omega))^\ctop ((\I_d - \lapl{\imbkernel}(0))^{-1})^\ctop \Omega_{\infty}^{-1} (\I_d - \lapl{\imbkernel}(\omega))^{-1}  (\I_d - \lapl{\imbkernel}(0))\, ,
\end{align*}
where we have introduced $\Omega_{\infty} := \lapl{\Omega}(0)$. Therefore, since $\ciperm$ is real, by \cref{prop:ci_kernel_func_phi} we get
$$
(\lapl{\cikernel^{'}}(\omega) + \cikernel(0)) \lapl{\Omega}(\omega) (\lapl{\cikernel^{'}}(\omega) + \cikernel(0))^{\ctop} = \ciperm \Omega_{\infty} \ciperm^\top \, .
$$
\end{proof}

We now prove \cref{prop:spectral_factor_kernel}. First, note that $\lapl{\Omega}$ admits a spectral factor. since it is positive-definite almost everywhere on the unit circle. Furthermore, by \cref{thm:titchmarsh} and \cref{ass:diff_0}, $\imbkernel$ is square-integrable and causal. Thus, \cref{lem:consistency_K} shows that $(\I_d - \lapl{\imbkernel})^{-1} \diag(\sqrt{\avgN})$ is a spectral factor of $\lapl{\Omega}$. Since there exists a spectral factor, the Paley-Wiener condition \cref{eq:paley-wiener} is necessarily satisfied.
\\ \\ 
As $\lapl{\Omega}$ satisfies the conditions of \cref{thm:mat-spectral-factor}, let $\mathcal{L}$ be any spectral factor of $\lapl{\Omega}$. Then, since the products of two functions of causal inverse Fourier transform also has a causal inverse Fourier transform, the inverse Fourier transform of $\lapl{\cikernel^{'}} \mathcal{L}$ is causal. Furthermore, since $\lapl{\cikernel^{'}}$ and $\mathcal{L}$ are both integrable functions, their product is square-integrable. Thus, the product satisfies the first condition of \cref{thm:titchmarsh} and therefore the product belongs to the Hardy space $\Hardy$. Therefore, by \cref{lem:consistency_K}, $\lapl{\cikernel^{'}} \mathcal{L}$ is a spectral factor of the positive-definite matrix $\ciperm \Omega_{\infty} \ciperm^\top$. It follows by the uniqueness of spectral factors up to a unitary matrix that there exists some unitary matrix $\mathcal{O}$ such that for almost all $\omega \in \R$
$$
\lapl{\cikernel^{'}}(\omega) = \mathcal{G} \mathcal{O} \mathcal{L}(\omega)^{-1} - \cikernel(0)\, ,
$$
where $\mathcal{G} \mathcal{G}^\top = \ciperm \Omega_{\infty} \ciperm^\top$. This concludes the proof.

\subsection{Proof of \cref{prop:imm_impact_matrix}}
\label{app:proof_imm_matrix}

The immediate cross-impact matrix $\cikernel(0)$ describes how trades push prices on very short time scales. As such, we intuitively expect that it must be constrained to prevent pair-trading arbitrage. In fact, we show in this section that it can be completely characterised. The first lemma shows that the immediate cross-impact matrix must be symmetric non-negative.

\begin{lemma}
\label{lemma:imm_impact_symm}
The immediate cross-impact matrix $\cikernel(0)$ is symmetric and non-negative.
\end{lemma}

\begin{proof}
This proof uses no-arbitrage arguments. It is inspired from \cite{Schneider2017Cross-impactNo-dynamic-arbitrage} but adapted here since assumptions are slightly different. We consider a deterministic trading strategy ending at time $T$ given by the function $f \colon [0,T] \to \R^{\nassets}$ which determines the buy and sell market orders according to \cref{def:trading_strategy}.  We re-write the cross-impact contributions as $\cikernel(t) = M + H(t)$, where $M$ is the immediate impact matrix and $H(0)=0$. From \cref{prop:ci_kernel_func_phi} we know that $H$ is continuous at zero. With these conventions, the average cost $C(f)$ of the trading strategy $f$ given in \cref{def:trading_strategy} is written as
$$
C(f) = \int_0^T f(t)^\top \int_0^t M f(s) \diff s \diff t + \int_0^T f(t)^\top \int_0^t H(t-s) f(s) \diff s \diff t =: C_i(f) + C_s(f),
$$
where we have split costs into two parts. The first, $C_i(f)$, represents immediate impact costs and the second, $C_s(f)$, represents the rest. Then
\begin{align*}
    C_i(f) =& \dfrac{1}{2} \sum_{i=1}^{\nassets} M_{ii} \normOne{f_i}^2 + \sum_{i \neq j}^{\nassets} M_{ij} \int_0^T \int_0^t f_i(t) f_j(s) \diff s \diff t.
\end{align*}
Similarly, for the other impact costs, we have
\begin{align*}
    C_s(f) = \sum_{i=1}^{\nassets} \int_0^T \int_0^t f_i(t) H_{ii}(t-s) f_i(s) \diff s \diff t + \sum_{i \neq j}^{\nassets}  \int_0^T \int_0^t f_i(t) H_{ij}(t-s) f_j(s) \diff s \diff t.
\end{align*}
We choose two distinct assets, Asset $a$ and Asset $b$ and consider, as in \cite{Schneider2017Cross-impactNo-dynamic-arbitrage}, a round-trip pair-trading strategy of the following form, where $v_p, v_q \in \R$:
$$
f_p(t) := \begin{cases}
v_p & \text{ for } 0 \leq t \leq T/3 \\
0 & \text{ for } T/3 \leq t \leq 2T/3 \\
-v_p & \text{ for } 2T/3 \leq t \leq T
\end{cases}, \hspace{1cm}
f_q(t) := \begin{cases}
v_q & \text{ for } 0 \leq t \leq T/3 \\
-v_q & \text{ for } T/3 \leq t \leq 2T/3 \\
0 & \text{ for } 2T/3 \leq t \leq T
\end{cases}.
$$
This strategy only trades Asset $p$ and Asset $q$, so that for all other Asset $i$, $f_{i} = 0$. This is a round-trip strategy since $\int_{0}^{T} f = 0$. Then, the immediate impact costs contribution is
\begin{align*}
    C_i(f) =& \dfrac{T^2}{18} (M_{pq} - M_{qp}) v_p v_q.
\end{align*}
For the other impact costs, we have
\begin{align*}
    C_s(f) &= \int_0^T \int_0^t f_a(t) H_{pp}(t-s) f_p(s) \diff s \diff t + \int_0^T \int_0^t f_q(t) H_{qq}(t-s) f_q(s) \diff s \diff t \\ 
    &+\int_0^T \int_0^t f_p(t) H_{pq}(t-s) f_q(s) \diff s \diff t + \int_0^T \int_0^t f_q(t) H_{qp}(t-s) f_q(s) \diff s \diff t.
\end{align*}
Therefore, since $H(0) = 0$ and $H$ is continous at $t=0$, for small enough execution times $T$ we have that for all $t \in [0,T]$, $\modulus{H_{ij}(t)} \leq \epsilon$ for all $(i,j) \in \{p,q\}^{2}$. Thus, combining both impact terms, we get
$$
\dfrac{C(f)}{T^2} \leq \dfrac{v_a v_b}{18}(M_{qp} - M_{pq}) + \epsilon.
$$
So, unless $M_{pq} = M_{qp}$, the volumes $v_p$ and $v_q$ can be chosen so that trading costs of this round trip strategy are negative. Thus, if $M_{pq} \neq M_{qp}$, arbitrages are possible. Therefore the immediate cross-impact matrix $M = \cikernel(0)$ is necessarily symmetric.
\\ \\
Since $\cikernel(0)$ is symmetric and $\cikernel$ is continuous by \cref{prop:ci_kernel_func_phi}, Lemma 2.8 of~\cite{Alfonsi2016MultivariateFunctions} implies that $\cikernel(0)$ is non-negative definite and hence non-negative. This completes the proof.
\end{proof}

The previous results show that the immediate cross-impact matrix is symmetric non-negative. By further using the price dynamics, we are able to relate it to the instantaneous covariance matrix of prices and order flows. 
\\ \\
We have
$$
\dd P_t = \cikernel(0) \diag(\avgvolume) (\dd N^{b}_t- \dd N^{a}_t) \, .
$$
Taking the predictable quadratic variation of the processes, we obtain
$$
\dd \covar{P, P}_t = \cikernel(0) \diag(\avgvolume^2) \diag({\lambda^a_t} + \lambda^b_t) \cikernel(0)^\top \dd t
$$
Taking expectations on both sides and writing the return covariance matrix $\Sigma_t \dd t :=  \E[ \dd \covar{P, P}_t]$ and the average intensity $\avgN_t := \E[\lambda^{b}_t] = \E[\lambda^{a}_t]$ we get
$$
\dfrac{1}{2} \Sigma_t = \cikernel(0) \diag(\avgvolume^2 \avgN_t) \cikernel(0)^\top \, .
$$
Since this holds for all $t$, and $\theta_t \underset{t \to \infty}{\to} \avgN = (\I_{\nassets} - \Phi^{a/a}) \mu + \Phi^{a/b} \mu$, it must hold as $t$ tends to infinity and we get
$$
\dfrac{1}{2}\Sigma = \cikernel(0) \diag (\avgN \avgvolume^2) \cikernel(0)^\top \, ,
$$
where $\Sigma = \underset{t \to \infty}{\lim}  \E[\dd \covar{P_t, P_t}]$, which is well-defined by passing to the limit in the above. Since, by \cref{lemma:imm_impact_symm}, $\cikernel(0)$ is symmetric non-negative and satisfies the above, it must be of the form (see for example Proposition 3 of~\cite{tomas2020build})
\begin{equation*}
\cikernel(0) = \dfrac{1}{\sqrt{2}} (\Lc_0^{-1})^\top \sqrt{\Lc_0^\top \Sigma \Lc_0} \Lc_0^{-1} \, ,
\end{equation*}
where $\Lc$ is given in the proposition. This completes the proof of \cref{prop:imm_impact_matrix}.

\subsection{Proof of \cref{prop:perm_impact_matrix}}
\label{app:proof_perm_matrix}

As in the proof of \cref{app:proof_imm_matrix}, we consider a trading strategy ending at time $T$ given by the function $f \colon [0,T] \to \R^{\nassets}$.  We re-write the cross-impact contributions as $\cikernel(t) = \ciperm + \tcikernel(t)$, where $\tcikernel(t) \underset{t \to \infty}{\to} 0$. Then, average cost $C(f)$ of the trading strategy $f$ given by \cref{def:trading_strategy} can be re-written as
$$
C(f) := \int_0^T f(t)^\top \int_0^t \ciperm f(s) \diff s \diff t + \int_0^T f(t)^\top \int_0^t \tcikernel(t-s) f(s) \diff s \diff t = C_p(f) + C_t(f) \, ,
$$
where we have split into permanent and temporary impact costs. Then
\begin{align*}
    C_p(f) =& \dfrac{1}{2} \sum_{i=1}^{\nassets} \ciperm_{ii} \normOne{f_i}^2 + \sum_{i \neq j}^{\nassets} \ciperm_{ij} \int_0^T \int_0^t f_i(t) f_j(s) \diff s \diff t.
\end{align*}
Similarly, for the temporary impact costs, we have
\begin{align*}
    C_t(f) = \sum_{i=1}^{\nassets} \int_0^T \int_0^t f_i(t) \tcikernel_{ii}(t-s) f_i(s) \diff s \diff t + \sum_{i \neq j}^{\nassets}  \int_0^T \int_0^t f_i(t) \tcikernel_{ij}(t-s) f_j(s) \diff s \diff t.
\end{align*}
Consider two distinct assets, Asset $a$ and Asset $b$. Consider, as in \cite{Schneider2017Cross-impactNo-dynamic-arbitrage}, a round-trip trading strategy  of the following form, here $v_p, v_q \in \R$:
$$
f_a(t) := \begin{cases}
v_p & \text{ for } 0 \leq t \leq T/3 \\
0 & \text{ for } T/3 \leq t \leq 2T/3 \\
-v_p & \text{ for } 2T/3 \leq t \leq T
\end{cases}, \hspace{1cm}
f_q(t) := \begin{cases}
v_q & \text{ for } 0 \leq t \leq T/3 \\
-v_q & \text{ for } T/3 \leq t \leq 2T/3 \\
0 & \text{ for } 2T/3 \leq t \leq T
\end{cases}.
$$
Then, the permanent impact costs contribution is of the form
\begin{align*}
    C_p(f) =& (\ciperm_{pq} - \ciperm_{qp}) \dfrac{T^2}{18} v_p v_q.
\end{align*}
Similarly, for the temporary impact costs, we have
\begin{align*}
    C_t(f) =& \int_0^T \int_0^t f_p(t) \tcikernel_{pp}(t-s) f_p(s) \diff s \diff t + \int_0^T \int_0^t f_q(t) \tcikernel_{qq}(t-s) f_q(s) \diff s \diff t \\ 
    &+ \int_0^T \int_0^t f_p(t) \tcikernel_{pq}(t-s) f_q(s) \diff s \diff t + \int_0^T \int_0^t f_q(t) \tcikernel_{qp}(t-s) f_p(s) \diff s \diff t.
\end{align*}
Therefore, since $\tcikernel$ is power-law with all exponents strictly below 1, we have
$$
\dfrac{C(f)}{T^2} \underset{T \to \infty}{=} \dfrac{v_p v_q}{18}(\ciperm_{qp} - \ciperm_{pq}) + o(1),
$$
so that if $\ciperm_{qp} \neq \ciperm_{pq}$, arbitrages are possible. Therefore $\ciperm$ is necessarily symmetric.
\\ \\
We now show that $\ciperm$ is non-negative. Let $\tau \in \R$ and $\eta \in \R^\nassets$ and consider a trading strategy $f$ which buys portfolio $\eta$ and waits $\tau$ units of time to sell it. Then, the cost of this trading strategy is
$$
C(f) = \eta^\top (\ciperm + \tcikernel(\tau)) \eta,
$$
and by no-arbitrage, $C(f) \geq 0$. Therefore, for all $\tau \in \R$ and $\eta \in \R^\nassets$, we have
$$
\eta^\top \ciperm \eta \geq - \eta^\top \tcikernel(\tau) \eta,
$$
so that using the fact that $\tcikernel(\tau) \underset{\tau \to \infty}{\to} 0$, $\ciperm$ is non-negative. This concludes the proof of \cref{prop:perm_impact_matrix}.

\subsection{Proof of \cref{prop:differentiability}}
\label{app:proof_differentiability}

To prove \cref{prop:differentiability}, we proceed in three steps. First, we use polarization to show that it suffices to prove the result for $\xi^\ctop \Zc \xi$ for any $\xi \in \C^{\nassets}$. Second, we use sufficient conditions for smoothness on characteristic functions. Finally, we show that these conditions are satisfied in our setting.
\\ \\
We begin with a polarization identity. Since $\Zc$ is a continuous positive definite function, Theorem 2.10 of~\cite{Alfonsi2016MultivariateFunctions} shows that for every $\xi \in \C^{\nassets}$ the continuous function $\Zc_{\xi} \colon t \mapsto \xi^\ctop \Zc(t) \xi$ is positive definite. Furthermore, for all $1 \leq a,b \leq \nassets$, the component $e_b^\top \Zc(t) e_a$ is equal to
$$
\frac{1}{2}(\Zc_{e_a + e_b}(t) - i \Zc_{e_a - i e_b}(t) - (1-i)\Zc_{e_a}(t) - (1-i)\Zc_{e_b}(t)) \, .
$$ 
Thus it suffices to prove the result for $\Zc_{\xi}$ for $\xi \in \C^{\nassets}$, which we show below.
\\ \\
For all $\xi \in \C^{\nassets}$, we introduce the measure $\Q_{\xi} := \dfrac{\xi^\ctop \M \xi}{\xi^\ctop \M(\R) \xi}$. For any $\xi \in \C^{\nassets}$, $\Q_{\xi}$ is indeed a probability measure since the matrix-valued measure $\M$ is non-negative definite and of finite total variation. Thus, the function $\Zc_{\xi}$ is the characteristic function associated to the probability measure $\Q_{\xi} = \dfrac{\xi^\ctop \M \xi}{\xi^\ctop \M(\R) \xi}$. We now use Theorem 2.3.1 of~\cite{lukacs1970characteristic} which gives a sufficient condition for the smoothness of a characteristic function. To apply the theorem, we must show that for all $\xi \in \C^{\nassets}$ we have
$$
\underset{t \to 0}{\lim \inf} \dfrac{\modulus{\Zc_{\xi}(2t) - \Zc_{\xi}(0) + \Zc_{\xi}(-2t) - \Zc_{\xi}(0)}}{4t^2} < \infty \, .
$$
Using the definition of $\Zc_{\xi}$, the above condition is equivalent to
$$
\underset{t \to 0}{\lim \inf} \dfrac{\modulus{\xi^\ctop \left( \cikernel(2t) - \cikernel(0) + \cikernel(2t)^\top - \cikernel(0)^\top \right) \xi}}{4t^2} < \infty \, .
$$
However, by \cref{ass:diff_0}, $\cikernel$ and its derivative are continuously differentiable at zero and \cref{prop:imm_impact_matrix} shows that $\cikernel(0)$ is symmetric, so that
$$
\dfrac{\cikernel(2t) - \cikernel(0) + \cikernel(2t)^\top - \cikernel(0)^\top}{4t^2} \underset{t \to 0}{=} \dfrac{1}{2t}(\cikernel^{'}(0) + \cikernel^{'}(0)^\top) + \cikernel^{''}(0) + \cikernel^{''}(0)^\top + o(1) \, .
$$
Therefore the condition will be satisfied and the proposition proven if $\cikernel^{'}(0) = - \cikernel^{'}(0)^\top$, that is if $\cikernel^{'}(0)$ is antisymetric. Then Theorem 2.3.1 of~\cite{lukacs1970characteristic} yields that $\Zc_{\xi}$ is twice differentiable and the integrals converge absolutely. To prove this result, we will use the no-arbitrage condition and the smoothness of the cross-impact kernel around zero.
\\ \\
As in the proofs of \cref{app:proof_perm_matrix,app:proof_imm_matrix}, we consider a trading strategy ending at time $T$ given by the function $f \colon [0,T] \to \R^{\nassets}$. We re-write the cross-impact contributions as $\cikernel(t) = \cikernel(0) + t \cikernel^{'}(0) + \Rc(t)$, where $\Rc$ is such that $\Rc(0)=0$, $\Rc$ is continuously differentiable at zero and $\Rc(t) \underset{t \to 0}{=} o(t)$. For convenience, we write $M(t) := \cikernel(0) + t \cikernel(0)$. Then the trading cost of the trading strategy is
$$
C(f) := \int_0^T f(t)^\top \int_0^t M(t-s) f(s) \diff s \diff t + \int_0^T f(t)^\top \int_0^t \Rc(t-s) f(s) \diff s \diff t = C_1(f) + C_2(f).
$$
Where we have split contributions. Consider two distinct assets, Asset $a$ and Asset $b$ and the round-trip trading strategy  of the following form, where $v_a, v_b \in \R$:
$$
f_a(t) := \begin{cases}
v_a & \text{ for } 0 \leq t \leq T/3 \\
0 & \text{ for } T/3 \leq t \leq 2T/3 \\
-v_a & \text{ for } 2T/3 \leq t \leq T
\end{cases}, \hspace{1cm}
f_b(t) := \begin{cases}
v_b & \text{ for } 0 \leq t \leq T/3 \\
-v_b & \text{ for } T/3 \leq t \leq 2T/3 \\
0 & \text{ for } 2T/3 \leq t \leq T
\end{cases}.
$$
This strategy only trades Asset $a$ and Asset $b$, i.e. for all other Asset $i$, $f_{i} = 0$. This is a round-trip strategy since $\int_{0}^{T} f = 0$. Then, computing contributions as in \cref{lemma:imm_impact_symm} and using the fact that $\cikernel(0)$ is symmetric, we obtain
\begin{align*}
    C_1(f) =& \dfrac{-5T^3}{162} v_a v_b (\cikernel^{'}(0)_{ab} + \cikernel^{'}(0)_{ba}).
\end{align*}
For the other contribution, we have
\begin{align*}
    C_2(f) &= \int_0^T \int_0^t f_a(t) \Rc_{aa}(t-s) f_a(s) \diff s \diff t + \int_0^T \int_0^t f_a(t) \Rc_{bb}(t-s) f_a(s) \diff s \diff t \\ 
    &+\int_0^T \int_0^t f_a(t) \Rc_{ab}(t-s) f_b(s) \diff s \diff t + \int_0^T \int_0^t f_b(t) \Rc_{ba}(t-s) f_a(s) \diff s \diff t.
\end{align*}
Therefore, since $\Rc(0) = 0$ and $\Rc(t) \underset{t \to 0}{=} o(t)$, for small enough execution times $T$ we have that for all $t \in [0,T]$, $\mid \Rc_{ij}(t) \mid \leq \epsilon t$ for all $(i,j) \in \{a,b\}^{2}$. Thus, we then have, where $c_1 > 0$ is a constant independent of $f$, $T$ and $\epsilon$
$$
C_2(f) \leq \epsilon c_1 T^3 \, .
$$
Thus, combining both terms we obtain
$$
\dfrac{C(f)}{T^3} \leq c_2 v_a v_b (\cikernel^{'}(0)_{ab} + \cikernel^{'}(0)_{ba}) + \epsilon,
$$
where $c_2 \neq 0$ is a constant independent of $f$, $T$ and $\epsilon$. Therefore, unless $\cikernel^{'}(0)_{ab} = -\cikernel^{'}(0)_{ba}$, the round-trip strategy yields negative costs and arbitrages are possible. Therefore $\cikernel^{'}(0)$ is necessarily antisymmetric. The proposition then follows.

\section{Calibration methodology details}

\label{sec:calibration_methodology}

This section details the calibration methodology. The objective is to calibrate the cross-impact kernels $\mcikernel$ and $\acikernel$ on empirical data. \cref{app:data_preparation} gives additional details on the dataset. \cref{app:estimation_empirical_observables} details the methodology for the estimation of the empirical observables $\Sigma$ and $\Omega$. \cref{app:estimation_spectral_factor} explains how spectral factors of the reduced covariance measure are computed. Finally, \cref{app:computation_kernels} explains the construction of both kernels.

\subsection{Data preparation and processing}
\label{app:data_preparation}

The data is processed according to the procedure outlined in \cite{tomas2020build}. For the reader's convenience, we recall some key elements here. The two instruments selected are the leading and third month E-Mini Futures. Prices and trades are gathered from anonymous trades and quotes data from the CME and prices are taken as the mid-price of the best bid and ask prices of each instrument. To avoid stationarity issues, we remove data outside the commonly traded hours of both instruments and the first and last 30 minutes of the trading period. Data ranges from January 2015 to December 2018.

\subsection{Estimation of empirical observables}
\label{app:estimation_empirical_observables}

Using the data outlined in \cref{app:data_preparation}, we detail here the estimation methodology for $\Sigma$ and $\Omega$. To simplify computations and reduce noise, we aggregate order flows and prices by bins of 1 seconds. Each bin contains the opening and closing price $p^o_t$ and $p^c_t$ as well as the total signed order flow $q_t$. Both of these random variables have zero mean. Using these conventions, we use naive statistical estimators to compute the price-covariance and the order flow auto-covariance, namely, for a daily timeseries of prices and signed order flows $\{p_t\}_{t=1}^{T}$ and $\{q_t\}_{t=1}^{T}$ at bins of one second:
\begin{align*}
    \Sigma &= \dfrac{1}{T-1} \sum_{t=1}^{T} (p^c_{t} - p^o_{t}) (p^c_{t} - p^o_{t})^\top \\
    \Omega(\tau) &= \dfrac{1}{T} \sum_{t=1}^{T} q_{t+\tau} q_t^\top \, .
\end{align*}
This procedure is averaged across days to obtain an empirical estimate of $\Sigma$ and $\Omega$ using approximately 800 days of available data. For stationary ergodic point processes, these averages do converge towards the theoretical price-covariance and reduced covariance measure by Proposition 8.3.1 of \cite{daley2008introduction}.

\subsection{Estimation of a spectral factor of $\Omega$}
\label{app:estimation_spectral_factor}

Once $\Sigma$ and $\Omega$ have been estimated, the cross-impact kernel $\mcikernel$ can be computed by solving numerically \cref{eq:spectral_factor_kernel} and setting the boundary conditions imposed by arbitrage-admissibility. To do so, we need to compute a spectral factor of $\Omega$.
\\ \\
To solve \cref{eq:spectral-factorization}, we use the SBR2 algorithm \cite{mcwhirter2007evd,wang2015multichannel}, a polynomial eigenvalue decomposition method, to compute a numerical approximation of a spectral factor of $\Omega$, $\mathcal{L}$. The implementation details follow the simple version of this algorithm (and not its subsequent improvements) outlined in \cref{eq:spectral-factorization}. The algorithm was tested on the examples provided in both papers and results were similar to those reported in the papers. To estimate the accuracy of the spectral decomposition, we computed the Frobenius norm of the error matrix $\dfrac{\normF{\Omega(z) - \mathcal{L}(z) \mathcal{L}(1/z)^\ctop}}{\normF{\Omega(z)}} \approx 6 \cdot 10^{-8}$. Thus, numerically, $\mathcal{L}$ is a good approximation for a spectral factor of $\Omega$.
\\ \\
A key property is that this spectral factor has the convenient form $\mathcal{L}(z) = D(z) H(z)$ where $H$ is a para-unitary polynomial matrix and $D$ is (close to) a diagonal polynomial matrix. Both $D$ and $H$ are outputs of the SBR2 algorithm. To check how close $D$ is to a diagonal matrix, we compute the Frobenius norm of its off-diagonal elements relative to its diagonal elements. The numerical results yield $10^{-15}$, which shows that the decomposition was successful. Once this spectral factor has been obtained, we need to compute the matrix polynomial $\mathcal{L}^{-1}$ to obtain the martingale-admissible kernel $\mcikernel$.

\subsection{Computation of the cross-impact kernels}
\label{app:computation_kernels}

Because of the structure of $\mathcal{L}(z)$, computing its inverse is straightforward. One the one hand, the inverse of the para-unitary polynomial matrix $H(z)$ is $H(1/z)^\ctop$. On the other hand, the inverse of $D(z)$ is obtained by taking the inverse of its diagonal elements, which is straightforward using pole decomposition. This allows us to compute numerically the polynomial matrix $\mathcal{L}^{-1}$ and, using \cref{eq:spectral_factor_kernel}, the martingale-admissible kernel $\mcikernel$.
\\ \\
The numerical computation of the poles of polynomials is done using the \texttt{residuez} function of the \texttt{signal} library of the \texttt{scipy} module \cite{2020SciPy-NMeth} of the \texttt{python} programming language \cite{python:ref}. All poles found were strictly inside the unit circle. However, they were numerically close to modulus one, which is consistent with the long-range auto-correlations of $\Omega$ reported in \cref{fig:omega}.
\\ \\
Once the martingale-admissible kernel $\mcikernel$ has been obtained, we derive the nsa-admissible kernel $\acikernel$ in the following manner. First, we compute the (numerical) Fourier transform of $\mcikernel$. The symmetric part of this Fourier transform is then modified so that all its eigenvalues are non-negative. Finally, this clipped Fourier transform is added to the asymmetric part of the Fourier transform of $\mcikernel$. The (numerical) inverse Fourier transform then yields the nsa-admissible kernel $\acikernel$.

\end{document}